\documentclass{revtex4}
\usepackage{amsmath}
\usepackage{amsfonts}
\usepackage{amssymb}
\usepackage{graphicx}
\usepackage{subfigure}
\newcommand{\rmd}{\text{d}}
\newcommand{\inpro}[2]{\left\langle #1 , #2 \right\rangle}
\newcommand{\norm}[1]{\Vert #1 \Vert}
\newcommand{\dfracd}[2]{\dfrac{\rmd #1}{\rmd #2}}
\newcommand{\dfracp}[2]{\dfrac{\partial #1}{\partial #2}}
\newcommand{\fracpp}[3]{\frac{\partial^{2} #1}{\partial #2 \partial #3}}
\newcommand{\up}[1]{\overline{#1}}
\newcommand{\TsM}{T^{\ast}M}
\newcommand{\til}[1]{\widetilde{#1}}
\newcommand{\gradH}{\text{grad}H}
\newcommand{\gradK}{\text{grad}K}

\newcommand{\Xbar}{
  \begin{picture}(10,10)
    \put(0,0){\text{\LARGE $x$}}
    \put(2,4){\line(1,0){7}}
  \end{picture}
}
\newcommand{\christoffel}[3]{
    \left\{ 
      \begin{array}{@{}c@{}c@{}}
        \multicolumn{2}{c}{#1} \\ #2 & #3
      \end{array} 
    \right\} }

\newcommand{\bR}{\boldsymbol{R}}
\newcommand{\bT}{\boldsymbol{T}}
\newcommand{\bU}{\boldsymbol{U}}
\newcommand{\bV}{\boldsymbol{V}}
\newcommand{\bW}{\boldsymbol{W}}
\newcommand{\bX}{\boldsymbol{X}}
\newcommand{\bY}{\boldsymbol{Y}}
\newcommand{\bZ}{\boldsymbol{Z}}

\newcommand{\bg}{\boldsymbol{g}}

\newcommand{\bxi}{\boldsymbol{\xi}}

\begin{document}
\draft

%\preprint{HEP/123-qed}

\title{Geometric Approach to Lyapunov Analysis in Hamiltonian Dynamics}

\author{YAMAGUCHI Y. Yoshiyuki 
  \footnote{e-mail: yyama@amp.i.kyoto-u.ac.jp}
  and IWAI Toshihiro
  \footnote{e-mail: iwai@amp.i.kyoto-u.ac.jp}}
\address{
  Department of Applied Mathematics and Physics,\\
  Kyoto University, Kyoto, 606-8501, Japan}
\date{\today}

%%%%%%%%%%%%%%%%%%%%%%%%%%%%%%%%%%%%%%%%%%%%%%%%%%%%%%%%%%%
% Abstract
%%%%%%%%%%%%%%%%%%%%%%%%%%%%%%%%%%%%%%%%%%%%%%%%%%%%%%%%%%%
\begin{abstract}
As is widely recognized in Lyapunov analysis, linearized Hamilton's
equations of motion have two marginal directions for which the
Lyapunov exponents vanish.  Those directions are the tangent one to a 
Hamiltonian flow and the gradient one of the Hamiltonian function.  
To separate out these two directions and to apply Lyapunov analysis 
effectively in directions for which Lyapunov exponents are not trivial, 
a geometric method is proposed for natural Hamiltonian systems, 
in particular.  
In this geometric method, Hamiltonian flows of a natural Hamiltonian system 
are regarded as geodesic flows on the cotangent bundle of a Riemannian
manifold with a suitable metric. Stability/instability of the geodesic
flows is then analyzed by linearized equations of motion which are related
to the Jacobi equations on the Riemannian manifold.  On some geometric
setting on the cotangent bundle, it is shown that along a geodesic
flow in question, there exist Lyapunov vectors such that two of them
are in the two marginal directions and the others orthogonal to the
marginal directions. It is also pointed out that Lyapunov vectors with
such properties can not be obtained in general by the usual method
which uses linearized Hamilton's equations of motion. Furthermore, it is
observed from numerical calculation for a model system that Lyapunov
exponents calculated in both methods, geometric and usual, coincide
with each other, independently of the choice of the methods.
\end{abstract}

\pacs{02.40.Ky,02.90.+p,05.45.Jn}

\maketitle

%%%%%%%%%%%%%%%%%%%%%%%%%%%%%%%%%%%%%%%%%%%%%%%%
% Section 1
%%%%%%%%%%%%%%%%%%%%%%%%%%%%%%%%%%%%%%%%%%%%%%%%
\section{Introduction}
\label{sec:introduction}

Natural Hamiltonian systems with many degrees of freedom have 
Hamiltonian functions of the form 
\begin{equation}
  \label{eq:natural-Hamiltonian}
  H(q,p) = \dfrac{1}{2} \sum_{i,j}^{N} \delta^{ij} p_{i} p_{j} + V(q). 
\end{equation}
In spite of the simple appearance, those Hamiltonian functions having 
appropriately chosen potential functions are used in a wide variety of 
physical sciences such as plasma physics, condensed matter physics, 
and celestial mechanics. 
However, the potential functions describe nonlinear interactions, 
in general, so that chaotic or highly unstable trajectories take place 
in respective phase spaces, as is widely recognized. 
The exponential instability of trajectories are measured in terms of 
Lyapunov exponents, which describe time-averaged properties of 
chaotic trajectories.  Further, in the study of directional deviations 
of chaotic trajectories, Lyapunov vectors will be of great use. 

The Lyapunov exponents and the Lyapunov vectors are defined through 
linearized Hamilton's equations of motion.  
For the Hamiltonian (\ref{eq:natural-Hamiltonian}), the linearized 
equations take the form
\begin{equation}
  \label{eq:linearized-Hamilton-eq}
  \dfracd{Q^{i}}{t} = P_{i},
  \quad
  \dfracd{P_{i}}{t} = - \sum_{j=1}^{N} 
  \fracpp{V}{q^{i}}{q^{j}}(q(t)) Q^{j},
  \quad
  i=1,\cdots,N,
\end{equation}
where $\bX=(Q^{1},\cdots,Q^{N},P_{1},\cdots,P_{N})$
is a $2N$-dimensional vector representing a deviation 
from a reference trajectory $(q(t),p(t))$ to a nearby trajectory.
The linearized equations (\ref{eq:linearized-Hamilton-eq}) have 
$2N$ linearly independent solutions, 
which we denote by $\bX_{a}(t),\ a=1,\cdots,2N$. 
The Lyapunov vectors $\bV_{a}(t)$ are then obtained by 
orthogonalizing these solutions on the Gram-Schmidt method: 
\begin{equation}
  \label{eq:Lyapunov-vectors}
  \bV_{a}(t) = \bX_{a}(t) 
  - \sum_{b=1}^{a-1}
  \dfrac{\inpro{\bX_{a}(t)}{\bV_{b}(t)}}
  {\inpro{\bV_{b}(t)}{\bV_{b}(t)}}
  \bV_{b}(t), 
  \qquad a=1,\cdots,2N,
\end{equation}
where $\inpro{\bX}{\bV}$ denotes the inner product of 
$\bX$ and $\bV$. 
The $a$-th Lyapunov exponent $\lambda_{a}$ is calculated as 
\begin{equation}
  \label{eq:Lyapunov-exponents}
  \lambda_{a} = \lim_{t\to\infty} \dfrac{1}{t} \ln 
  \dfrac{\norm{\bV_{a}(t)}}{\norm{\bV_{a}(0)}}. 
\end{equation}
It is to be noted that the values of the Lyapunov exponents are known 
to be independent of the choice of initial values of the Lyapunov 
vectors except for vanishing Lebesgue measure \cite{oseledec-68,benettin-76},
and that the exponents are ordered as
$\lambda_{1}\geq\lambda_{2}\geq\cdots\geq\lambda_{2N}$.

Since the Lyapunov exponents are time-averaged quantities, 
they are suitable for the study of statistic properties of Hamiltonian 
systems.  For example, phase transitions are investigated by the use of 
Lyapunov exponents. 
In fact, the second-order phase transition \cite{firpo-98} and 
the Kosterlitz-Thouless transition \cite{butera-87} are characterized 
by the discontinuity in the largest Lyapunov exponents 
and by a sudden change in the gradient of the largest Lyapunov exponent 
against energy, 
respectively. 
Further, the sum of all positive Lyapunov exponents, which is 
also viewed as a function of energy, 
is used in the discussion of a dynamical phase transition 
\cite{mutschke-93}, 
according to which trajectory's phase transition from nearly-integrable 
behavior to chaotic behavior occurs in an energy region 
in which the sum of positive exponents breaks 
into a rapid increase against energy. 
In contrast with this, the Lyapunov vectors are expected to be useful in 
studying dynamical behavior of chaotic trajectories, 
since they serve as time series
\cite{konishi-92,sasa-00}.

Suppose that a reference trajectory is given in a phase space. 
Then, according to (\ref{eq:Lyapunov-vectors}), one can form 
$2N$ linearly independent Lyapunov vectors from solutions to 
the linearized equations of motion along the reference trajectory. 
However, two of the Lyapunov vectors which are associated with 
Lyapunov exponents, $\lambda_{N}$ and $\lambda_{N+1}$, 
are considered as marginal, since  $\lambda_{N}$ and $\lambda_{N+1}$ 
should vanish, as is widely recognized.  
One of those two Lyapunov vectors is the tangent vector to the trajectory, 
$\bX_{\!H}$, and the other the gradient vector of Hamiltonian function, 
$\gradH$.  
We may interpret these vectors as follows: 
The displacement in the direction $\bX_{\!H}$ is regarded just 
as a certain time displacement in the reference trajectory, 
and the displacement in the direction $\gradH$ will give rise to 
a transfer to a nearby trajectory with an energy value different from 
that of the reference trajectory.  
In view of this, in order to analyze the instability of trajectories, 
we are allowed to require that the two directions pointed by the 
vectors $\bX_{\!H}$ and $\gradH$ be separated out from the 
other $2N-2$ directions.  
Put another way, the requirement means that a Lyapunov vector 
which is orthogonal to the plane spanned by $\bX_{\!H}$ and $\gradH$ 
at an initial instant has to be orthogonal to the plane spanned by 
$\bX_{\!H}$ and $\gradH$ at every instant. 
If the requirement is fulfilled, we will be able to discuss 
the instability of trajectories without influence of the two marginal 
directions. 

Unfortunately, the usual method of Lyapunov analysis on the basis of 
the equations (\ref{eq:linearized-Hamilton-eq}) does not satisfy the
requirement in general.  This is because 
for any solution $\bX(t)$ to (\ref{eq:linearized-Hamilton-eq})
one has 
\begin{equation}
  \label{eq:inpro-X-gradH}
  \dfracd{}{t} \inpro{\bX}{\gradH} = 0, 
\end{equation}
so that one obtains $\inpro{\bX}{\gradH}=0$ at any instant 
if $\inpro{\bX}{\gradH}|_{t=0}=0$ at an initial instant, 
but, in general, by no means one can make $\inpro{\bX}{\bX_{\!H}}$ 
vanish at any instant, so that even the first Lyapunov vector,
$\bV_{1}$, can not be made orthogonal to the plane spanned by $\bX_{\!H}$ 
and $\gradH$ at every instant.  

A way to construct Lyapunov vectors which satisfy the above-stated 
requirement is to adopt linearized equations of different type  
from the usual one (\ref{eq:linearized-Hamilton-eq}).  
To take a geometric approach to Hamilton's equations of motion is 
a step toward finding such Lyapunov vectors. 
As for the geometric approach, it is known that if the total energy of 
the natural dynamical system is fixed at $E$, Newton's equations of motion 
can be equivalently expressed as geodesic equations on a Riemannian manifold 
$(M, g_{ij})$, where $M$ is a subspace of the configuration space 
$\bR^{N}$ defined by $M=\{q\in \bR^{N}|\,E-V(q)>0\}$ and $g_{ij}$ is 
the Jacobi metric defined  by $g_{ij}(q)=2(E-V(q))\delta_{ij}$. 
Then the linearized equations of the geodesic equations are given 
by the Jacobi equations of the form
\begin{equation}
  \label{eq:Jacobi-eq}
  \dfrac{\rmd^{2}X^{i}}{\rmd s^{2}} 
  + \sum_{j,k,\ell=1}^{N}
  R_{jk\ell}^{\hspace*{1.2em}i} X^{j} \dfracd{q^{k}}{s}
  \dfracd{q^{\ell}}{s} = 0,
  \quad i=1,\cdots,N,
\end{equation}
where $R_{jk\ell}^{\hspace*{1.2em}i}$ are the components of the 
Riemann curvature tensor, and $s$ is the arc length defined as
\begin{equation}
  \label{eq:arc-length-s}
  \rmd s^{2} = \sum_{i,j=1}^N g_{ij}(q) \rmd q^{i}\rmd q^{j}
  = \sum_{i,j=1}^N g_{ij}(q) \dfracd{q^{i}}{t} \dfracd{q^{j}}{t} \rmd t^{2}.
\end{equation}
On the other hand, the linearized Newton's equations are put 
in the form  
\begin{equation}
  \label{eq:linearized-Newton-eq}
  \dfrac{\rmd^{2}Q^{i}}{\rmd t^{2}}
  + \sum_{j=1}^{N} \fracpp{V}{q^{i}}{q^{j}} Q^{j} = 0,
  \quad
  i=1,\cdots,N.
\end{equation}
Equations (\ref{eq:Jacobi-eq}) and (\ref{eq:linearized-Newton-eq}) are 
not transformed to each other through the parameter transformation 
(\ref{eq:arc-length-s}), while Newton's equations of motion and geodesic 
equations for the Jacobi metric are transformed to each other. 
This geometric method has been introduced in the estimation of 
the largest Lyapunov exponent $\lambda_{1}$ with the aid of 
statistical mechanics  
%\cite{pettini-93,casetti-93,casetti-00}.  
\cite{pettini-93,casetti-95,casetti-96,casetti-00}.
They studied instability of geodesics through the Jacobi equation, 
a second-order differential equation, while the Lyapunov analysis 
needs first-order differential equations. 

The geometric approach we will take in this article is to be made on 
the cotangent bundle $\TsM$ of the Riemannian manifold $M$ 
in order to find first-order differential equations associated with 
(\ref{eq:Jacobi-eq}) and thereby to construct Lyapunov vectors 
which satisfy the above-stated requirements. 
We will first work with generic linearized Hamilton's equations 
of motion on $\TsM$, and then specialize the resultant equations 
to linearized Hamilton's equations for geodesic flows on $\TsM$, 
which will be found to project to the Jacobi equations on $M$.  
Further, we will introduce a lifted metric on the cotangent 
bundle $\TsM$ to make it possible to discuss the orthogonality of 
vector fields on $\TsM$. The lifted metric may be called the Sasaki metric. 
On this setting, we will be able to find Lyapunov vectors satisfying 
the above-stated requirements along any geodesic flow on $\TsM$. 
Put in detail, it will be shown that along any geodesic flow on $\TsM$, 
there exist Lyapunov vectors such that those associated with the 
vanishing Lyapunov exponents $\lambda_N$ and $\lambda_{N+1}$ are 
$\bX_{\!H}$ and $\gradH$, respectively, and the other $2N-2$ Lyapunov 
vectors are all orthogonal to the plane spanned by $\bX_{\!H}$ and 
$\gradH$ at each point of the geodesic flow.  

This article is organized as follows:  
Section \ref{sec:geometry} contains a brief review of geodesics and
Jacobi fields and, in particular, of the Jacobi metric,
whose geodesics are equivalent to trajectories of the natural
dynamical system with a fixed total energy. 
In Sec.\ref{sec:geometry} and succeeding sections, 
Einstein's summation convention is adopted,
and we choose to denote by $(x^{i})$ local coordinates on a general
$m$-dimensional Riemannian manifold, 
and by $(q^{i})$ the Cartesian coordinates on $\bR^{N}$.
Section \ref{sec:cotangent-bundle} is concerned with geodesic 
flows on the cotangent bundle $\TsM$, 
which project to geodesics on $M$.
To describe geodesic flows in a more geometric way, we introduce an 
adapted frame and a lifted Riemannian metric on $\TsM$. 
Linearized Hamilton's equations of motion are discussed 
in Section \ref{sec:lyapunov-analysis}, and it will be shown that 
there exist Lyapunov vectors which satisfy the above-stated
requirement along a geodesic flow on $\TsM$. 
Section \ref{sec:calculation} is for numerical calculations for 
a model system with three degrees of freedom. 
Lyapunov vectors and Lyapunov exponents are calculated numerically 
for the model system in both geometric and usual methods to 
compare the respective results.  
It will be shown that Lyapunov exponents calculated in respective 
methods coincide with each other, independently of the choice of 
methods.  
Section \ref{sec:summary} is devoted to concluding remarks.
Appendices are attached in which related topics on 
geometry of cotangent bundles and a symplectic implicit Runge-Kutta 
method for numerical integration are reviewed. In particular, 
lifting vector fields on $M$ to $\TsM$ and the Levi-Civita connection 
with respect to the lifted metric on $\TsM$ are discussed.

%%%%%%%%%%%%%%%%%%%%%%%%%%%%%%%%%%%%%%%%%%%%%%%%
% Section 2
%%%%%%%%%%%%%%%%%%%%%%%%%%%%%%%%%%%%%%%%%%%%%%%%
\section{Geodesics and Jacobi Fields}
\label{sec:geometry}

\subsection{Jacobi equations}

Let $(M,\bg)$ be an $m$-dimensional Riemannian manifold 
with metric $\bg$.
The metric induces the Levi-Civita connection $\nabla$ on $M$;
for vector fields $\bY,\ \bZ\in \Xbar(M)$,
$\Xbar(M)$ denoting the set of vector fields on $M$,
the covariant derivative $\nabla_{\bY} \bZ$ is defined,
in terms of local coordinates $(x^{1},\cdots,x^{m})$, to be
\begin{displaymath}
  \nabla_{\bY} \bZ
  = Y^{j} \left[
    \dfracp{Z^{i}}{x^{j}} + \Gamma^{i}_{jk} Z^{k}
  \right] \dfracp{}{x^{i}}, 
\end{displaymath}
where $(Y^{i})$ and $(Z^{i})$ are components of $\bY$ and
$\bZ$, respectively, the Christoffel symbols $\Gamma^{i}_{jk}$ are
defined as
\begin{displaymath}
  \Gamma^{i}_{jk} = \frac{1}{2} g^{i\ell} \left(
    \dfracp{g_{\ell k}}{x^{j}}
    + \dfracp{g_{j \ell }}{x^{k}}
    - \dfracp{g_{jk}}{x^{\ell}}
  \right) 
\end{displaymath}
with components of the metric
\begin{displaymath}
  g_{ij} = \bg \left( \dfracp{}{x^{i}}, \dfracp{}{x^{j}} \right),
  \qquad
  g_{ij} g^{jk} = \delta_{i}^{k} .
\end{displaymath}

For a geodesic $c(s)$ with $s$ the arc length parameter,
the tangent vector $\bxi$ to the geodesic satisfies the geodesic
equation
\begin{equation}
  \label{eq:geodesic-eq}
  \nabla_{\!\bxi} \bxi 
  = \left [ 
    \dfracd{\xi^{i}}{s} + \Gamma^{i}_{jk} \xi^{j} \xi^{k}
  \right] \left.\dfracp{}{x^{i}}\right|_{c(s)}
  = 0,
\end{equation}
where
\begin{displaymath}
  \bxi = \left. 
    \dfracd{x^{i}}{s} \dfracp{}{x^{i}} \right|_{c(s)}
\end{displaymath}
with
\begin{equation}
  \label{eq:s-t}
  \rmd s^{2} = g_{ij} \rmd x^{i} \rmd x^{j} .
\end{equation}

We are interested in stability/instability of geodesics.
To this end, we consider a congruence of geodesics which looks 
like a fluid whose flow lines are geodesics with the $c(s)$ 
as a member of them. 
Then we may consider that the tangent vector $\bxi$ to $c(s)$ is 
extended to be a vector field defined in a neighborhood of 
the original geodesic $c(s)$. 
We may also assume that there exists a vector field $\bY$ satisfying  
the condition 
\begin{equation}
  \label{eq:commutes}
  [\bxi,\bY]=0  
\end{equation}
in the same domain as that for $\bxi$.  
The condition (\ref{eq:commutes}) means that a geodesic with $\bxi$ 
its tangent vector is carried congruently to another infinitesimally 
nearby geodesic by the infinitesimal transformation $\bY$. 
Thus $\bY$ is viewed as a deviation of geodesics. 
The vector field $\bxi$ may have singularity at which $\bxi$ is not 
defined uniquely, and $\bY$ may vanish there. 
With this in mind, we operate $\nabla_{\!\bxi} \bxi=0$ 
with $\nabla_{\bY}$ and use the definition of the Riemann curvature tensor 
and of the torsion, which vanishes identically, 
to obtain the Jacobi equation
\begin{equation}
  \label{epppq:Jacobi-eq}
  \nabla_{\!\bxi} \nabla_{\!\bxi} \bY 
  + \bR( \bY, \bxi ) \bxi = 0.
\end{equation}
Here, as is well known, the torsion tensor and the Riemann 
curvature tensor are defined, respectively, to be 
\begin{displaymath}
  \begin{split}
    \bT(\bY,\bZ) 
    & = \nabla_{\bY} \bZ - \nabla_{\bZ} \bY 
    - [\bY,\bZ],\\[2mm]
    \bR(\bY,\bZ) \bW 
    & = \nabla_{\bY} \nabla_{\bZ} \bW 
    - \nabla_{\bZ} \nabla_{\bY} \bW
    - \nabla_{[\bY,\bZ]} \bW,     
  \end{split}
\end{displaymath}
where $\bY,\bZ,\bW\in\Xbar(M)$, and the Riemann curvature tensor 
has symmetries such as
\begin{equation}
  \label{eq:R-symmmetry}
  \begin{split}
    \bg( \bR (\bY,\bZ) \bW, \bU)
    & = - \bg( \bR (\bZ,\bY) \bW, \bU) \\[2mm]
    & = - \bg( \bR (\bY,\bZ) \bU, \bW)
    = \bg( \bR (\bW, \bU) \bY,\bZ),
  \end{split}
\end{equation}
where $\bY,\bZ,\bW,\bU\in\Xbar(M)$.
Local components of the Riemann curvature tensor are expressed as 
\begin{displaymath}
  \begin{split}
    R_{ijk\ell} 
    & = R_{ijk}^{\hspace*{1.2em}m} g_{m\ell}
    = \bg\left( 
      \bR\left(  \dfracp{}{x^{i}},\dfracp{}{x^{j}} \right)
      \dfracp{}{x^{k}}, \dfracp{}{x^{\ell}} 
    \right), \\
    R_{ijk}^{\hspace*{1.2em}\ell}
    & = \dfracp{\Gamma^{\ell}_{jk}}{x^{i}} 
    - \dfracp{\Gamma^{\ell}_{ik}}{x^{j}}
    + \Gamma^{\ell}_{im} \Gamma^{m}_{jk}
    - \Gamma^{\ell}_{jm} \Gamma^{m}_{ik} .
  \end{split}
\end{displaymath}

In the next subsection, we will give an example of Riemannian 
metrics whose geodesics are equivalent to trajectories of 
Newton's equations of motion
\begin{equation}
  \label{eq:Newton-eq}
  \dfrac{\rmd^{2} q^{i}}{\rmd t^{2}} + \dfracp{V}{q^{i}} = 0,
  \quad
  i=1,\cdots,N.
\end{equation}
Then, in order to analyze stability/instability of 
trajectories of the natural Hamiltonian system, we can deal with 
the Jacobi equation, a linearization of the geodesic equation.  
However, the Jacobi equations in their original form are not 
suitable to Lyapunov analysis. 

\subsection{Geodesics for the Jacobi metric}
\label{sec:jacobi-metric}

Consider equations of motion, Eq.(\ref{eq:Newton-eq}), on $\bR^{N}$, 
which we call a natural dynamical system. 
Let $M_{J}$ be an open submanifold of the configuration space 
$\bR^{N}$, which is defined to be 
\begin{equation}
  \label{eq:Jacobi-mfd}
  M_{J} = \{ q\in \bR^N | V(q) < E \}. 
\end{equation}
As is well known, if energy is fixed at $E$, almost all trajectories
are confined in $M_J$ when $N\geq 2$. 
On the other hand, the Jacobi metric $\bg_{J}$ is defined, in $M_J$,
to be
\begin{equation}
  \label{eq:Jacobi-metric}
  (g_{J})_{ij} = 2 [ E - V(q) ] \delta_{ij} .
\end{equation}
According to Maupertuis's Principle of Least Action, 
an extremal of the action, the integral of the kinetic energy 
along possible paths, provides an actual trajectory 
of total energy $E$. 
This principle can also be stated as follows: 
An extremal of the variational problem of lengths of paths 
with respect to the Jacobi metric provides 
an actual trajectory of the total energy $E$ \cite{whittaker-37,ong-75}. 
From Eq.(\ref{eq:s-t}) along with the Jacobi metric $\bg_J$, 
the arc length parameter $s$ is shown to be related to the time
parameter $t$ by
\begin{equation}
  \label{eq:s-t-Jacobi}
  \rmd s^{2} = 4 [ E - V(q) ]^{2} \rmd t^{2},
\end{equation}
and the tangent vector to a geodesic is always unity accordingly:
\begin{displaymath}
  \bg_J(\bxi,\bxi) = g_{ij} \xi^{i} \xi^{j}
  = 2(E-V) \delta_{ij} \dfracd{q^{i}}{s} \dfracd{q^{j}}{s}
  = [2(E-V)]^{2} \left( \dfracd{t}{s} \right)^{2} = 1.
\end{displaymath}

Since the Christoffel symbols for the Jacobi metric (\ref{eq:Jacobi-metric}) 
are given by 
\begin{equation}
  \label{eq:Gamma-J}
  \Gamma^{i}_{jk}=\frac{-1}{2(E-V)}\left[
      \dfracp{V}{q^j}\delta^i_k +\dfracp{V}{q^k}\delta^i_j 
        -\dfracp{V}{q^i}\delta_{jk}\right],
\end{equation}
the geodesic equations for the Jacobi metric are expressed as
\begin{displaymath}
  \dfrac{\rmd^{2} q^{i}}{\rmd s^{2}}
  - \dfrac{1}{E-V} 
    \dfracp{V}{q^{j}}
    \dfracd{q^{j}}{s} \dfracd{q^{i}}{s}
    + \frac{1}{4(E-V)^2}\dfracp{V}{q^{i}} 
  = 0 ,
\end{displaymath}
which prove to be equivalent to Newton's equations of motion
(\ref{eq:Newton-eq}) on account of (\ref{eq:s-t-Jacobi}).
However, the Jacobi equations (\ref{eq:Jacobi-eq}) with the curvature
tensor for the Jacobi metric are not brought into the same equations
as (\ref{eq:linearized-Newton-eq}), 
a linearization of Newton's equations of motion, in general.
Components of the curvature tensor for $\bg_J$ are indeed put 
in the form \cite{ong-75,casetti-00}
\begin{displaymath}
  R_{ijk\ell} = C_{i\ell} \delta_{jk} + C_{jk} \delta_{i\ell}
    - C_{ik} \delta_{j\ell} - C_{j\ell} \delta_{ik},
\end{displaymath}
where
\begin{displaymath}
  C_{ij} = \fracpp{V}{q^{i}}{q^{j}} 
    + \dfrac{3}{2(E-V)} \dfracp{V}{q^{i}} \dfracp{V}{q^{j}} 
    - \dfrac{1}{4(E-V)} \delta^{k\ell} \dfracp{V}{q^{k}} \dfracp{V}{q^{\ell}}
    \delta_{ij}. 
\end{displaymath}
%

%%%%%%%%%%%%%%%%%%%%%%%%%%%%%%%%%%%%%%%%%%%%%%%%
% Section 3
%%%%%%%%%%%%%%%%%%%%%%%%%%%%%%%%%%%%%%%%%%%%%%%%
\section{Geodesic flows on Cotangent Bundles}
\label{sec:cotangent-bundle}

In the previous section, we have mentioned that trajectories of 
a natural dynamical system with a fixed energy can be regarded 
as geodesics on a suitable Riemannian manifold, and 
that stability/instability of the
trajectories are analyzed through the Jacobi equation,
a linearization of the geodesic equation.
However, the Jacobi equation is a second-order differential equation,
while Lyapunov analysis is applied to first-order differential
equations.
We hence need a first-order differential equation associated 
with the Jacobi equation in order to apply Lyapunov analysis. 
To find such a first-order differential equation,
we are working on the cotangent bundle $\TsM$ of a Riemannian 
manifold $M$ along with some geometric setting-ups on $T^*M$.  
Related topics on $\TsM$ will be described in Appendix
\ref{sec:appendixA}.
\par
At first, let us be reminded of a minimum on cotangent bundles. 
Let $M$ be an $m$-dimensional  Riemannian manifold endowed with 
Riemannian metric $\bg=g_{ij}\rmd x^{i} \otimes \rmd x^{j}$, 
and $T^*M$ the cotangent bundle of $M$ with the projection 
$\pi:\,T^*M \to M$. 
Let $(x^{i})$ and $(x^{i},p_{i})$ be local coordinates in an 
open subset $U\subset M$ and in $\pi^{-1}(U)$, respectively. 
Further, let $(\up{x}^{i},\up{p}_{i})$ be another local coordinates  
in $\pi^{-1}(\overline{U})$ with $\pi^{-1}(U)\cap\pi^{-1}(\overline{U})
\neq \emptyset$.  Then one has the coordinate transformation 
in the intersection $\pi^{-1}(U)\cap\pi^{-1}(\overline{U})$,
\begin{equation}
  \label{eq:coordinates-transf}
  \up{x}^{i} = \up{x}^{i}(x), \quad
  \up{p}_{i} = \dfracp{x^{j}}{\up{x}^{i}} p_{j}.
\end{equation}

\subsection{Geodesic flows}
\label{sec:hamiltonian-flow}

We recall that Newton's equations of motion have been already 
``geometrized" so as to be geodesic equations on a suitable Riemannian 
manifold, so that further external force doesn't need to be taken 
into account anymore.  In other words, we have only to consider a free 
particle motion on $M$.  In the Hamiltonian formalism, 
the Hamiltonian we then have to study should be given, 
on $T^*M$, by 
\begin{equation}
  \label{eq:Hgeo}
  K = \dfrac{1}{2} g^{ij}(x) p_{i} p_{j}, 
\end{equation}
where $(g^{ij}):=(g_{ij})^{-1}$. 
Hamilton's equations of motion for $K$ are then put in the form 
\begin{equation}
  \label{eq:Hamilton-eq}
  \begin{split}
    \dfracd{x^{i}}{s} & = \dfracp{K}{p_{i}} = g^{ij} p_{j},\\
    \dfracd{p_{i}}{s} & = - \dfracp{K}{x^{i}}
    = g^{kj} \Gamma^{\ell}_{ji} p_{k} p_{\ell},
  \end{split}
\end{equation}
where use has been made of the equality 
\begin{displaymath}
  - \dfracp{g^{k\ell}}{x^{i}} 
  = g^{kj} \Gamma^{\ell}_{ji} + g^{j\ell} \Gamma^{k}_{ji}. 
\end{displaymath}
It is an easy matter to show that Eq.(\ref{eq:Hamilton-eq}) 
projects to geodesic equation on $M$.  
In fact, put together, differentiation of the first 
equation of (\ref{eq:Hamilton-eq}) with respect to $s$ and 
the second equation of (\ref{eq:Hamilton-eq}) 
along with the above equality 
provide geodesic equations. 
As is well known, Eq.(\ref{eq:Hamilton-eq}) is associated 
with the Hamiltonian vector field $\bX_{\!K}$ given by 
\begin{equation}
  \label{eq:Hamiltonian-flow-geo}
  \bX_{\!K} = \dfracp{K}{p_{i}} \dfracp{}{x^{i}}
  - \dfracp{K}{x^{i}} \dfracp{}{p_{i}}=
  g^{ij} p_{j}\dfracp{}{x^{i}}+
  g^{kj} \Gamma^{\ell}_{ji} p_{k} p_{\ell}\dfracp{}{p_{i}}.
\end{equation}
Integral curves of (\ref{eq:Hamiltonian-flow-geo}) are called 
geodesic flows.  We note here that the nomenclature ``geodesic 
flows" are usually assigned to the corresponding flows on the 
tangent bundle, but we use the word for convenience' sake. 
\par
It is worth noting here that how geodesic flows on $\TsM$ 
project to geodesics on $M$. 
Let $P(s)=(x(s),p(s))$ be a geodesic flow with an initial value 
$(x(0),p(0))=(a,b)$ with $g^{ij}b_ib_j=1$.  Define a tangent vector 
$v$ to $M$ at $a=\pi(P(0))$ by $v^i=g^{ij}b_j$.  
Then the projection $x(s)=\pi(P(s))$ is a geodesic with the initial 
value $(x(0),\dot x(0))=(a,v)$. Varying $b\in T^{*}_{a}M$ with 
$g^{ij}b_ib_j=1$ but fixing $a$, we obtain an $(m-1)$-parameter 
family of geodesic flows on $\TsM$, which projects to an $(m-1)$-parameter 
family of geodesics passing the point $a$ of $M$. 
Furthermore, the vector field $\bX_{\!K}$ projects to the tangent vector field 
$\bxi$ to this family of geodesics $M$.  
However, $\bxi$ has singularity only at $a\in M$ in a neighborhood of $a$. 
We have to note that if all geodesic flows on $T^*M$ project to geodesics 
$M$, those geodesics may not define such a tangent vector field uniquely. 
If there is another $(m-1)$-parameter family of geodesic flows on $T^*M$, 
it may project to another $(m-1)$-parameter family of geodesics on $M$ 
along with a tangent vector field like $\bxi$.

\subsection{Adapted frames}
\label{sec:adapted-frame}

To describe geodesic flows in a more geometric way, we introduce 
an adapted frame and an adapted coframe on $\pi^{-1}(U)\subset \TsM$ 
by the use of the Christoffel symbols $\Gamma^i_{jk}$ on $M$. 
The adapted frame and coframe are defined, in $\pi^{-1}(U)$, to be 
\begin{equation}
  \label{eq:adapted-frame}
  D_{i} = \dfracp{}{x^{i}} + p_{k} \Gamma^{k}_{ij} \dfracp{}{p_{j}
},
  \qquad
  D_{\up{i}} = \dfracp{}{p_{i}},
\end{equation}
and 
\begin{equation}
  \label{eq:adapted-form}
  \theta^{i} = \rmd x^{i}, \qquad
  \theta^{\up{i}} = \rmd p_{i} - p_{k} \Gamma^{k}_{ij} \rmd x^{j},
\end{equation}
respectively, where $\up{i}=i+m$.  
These frames are dual to each other, {\it i.e.}, satisfy
\begin{displaymath}
  \theta^{i}(D_{j}) = \delta^{i}_{j},
  \qquad \theta^{i}(D_{\up{j}}) = 0,
  \qquad \theta^{\up{i}}(D_{j})  = 0,
  \qquad \theta^{\up{i}}(D_{\up{j}}) = \delta_{i}^{j}.
\end{displaymath}
If there is another adapted frame $\{ \up{D}_{i}, \up{D}_{\up{i}} \}$
in an open set $\pi^{-1}(\up{U})$ and if the intersection 
$\pi^{-1}(\up{U})\cap \pi^{-1}(U)$ is non-empty, 
then from (\ref{eq:coordinates-transf}) it follows that 
the adapted frames are subject to the transformation 
\begin{equation}
  \label{eq:f-transf}
   \up{D}_{i}=\dfracp{x^{j}}{\up{x}^{i}} D_{j}, \quad 
   \up{D}_{\up{i}}=\dfracp{\up{x}^{i}}{x^{j}} D_{\up{j}}.
\end{equation}
For adapted coframes, an analogous transformation holds as well. 
\par
The transformation (\ref{eq:f-transf}) implies that 
$D_i,\,i=1,\cdots,m$, and $D_{\up{i}},\,\up{i}=m+1,\cdots,2m$, 
define, respectively, subspaces $H_{P}$ and $V_{P}$ of the tangent
space $T_P T^*M$ at each point $P\in T^*M$ independently of 
the choice of adapted frames. 
Thus one obtains a direct sum decomposition of the tangent space 
to $\TsM$ at each point $P\in\TsM$;
\begin{equation}
  \label{eq:decomp}
    T_{P}\TsM = H_{P} \oplus V_{P}.  
\end{equation}
The subspaces $H_{P}$ and $V_{P}$ are called the horizontal and 
the vertical subspace of $T_{P}\TsM$, respectively. 
We notice here that $H_{P}$ and $T_{\pi(P)}M$ are isomorphic as 
vector spaces. 
Note further that the transformation rule for the standard frame
$\{\partial/\partial x^{i},\partial/\partial p_{i}\}$ is mixed up, 
so that one cannot define a subspace, say, 
$\mbox{span}\{\partial/\partial x^{i}\}$ independently of the choice 
of natural frames.  
See \cite{yano-73} for adapted frames on the tangent bundle $TM$. 
\par
In terms of the adapted frame, the Hamiltonian vector field $\bX_{\!K}$
becomes expressed as  
\begin{equation}
  \label{eq:geod-spray}
  \bX_{\!K} = D_{\up{i}}(K) D_{i} - D_{i}(K) D_{\up{i}}
  = g^{ij} p_{j} D_{i}, 
\end{equation}
which shows that $\bX_{\!K}$ is a horizontal vector field and further 
that geodesic flows are horizontal curves in the sense that the tangent 
vectors to them are always horizontal.

\subsection{The Sasaki metric}
\label{sec:lift-metric}

As is already seen, the tangent space to $\TsM$ at each point 
of $\TsM$ is decomposed into a direct sum.  
We can define a metric on $\TsM$ so that the decomposition may be  
orthogonal direct sum.  One of such metrics is the Sasaki metric, 
which is a lifted metric $\til{\bg}$ given by 
\begin{equation}
  \label{eq:lift-metric}
  \til{\bg} = g_{ij} \theta^{i} \otimes \theta^{j} 
  + g^{ij} \theta^{\up{i}} \otimes \theta^{\up{j}}.
\end{equation}
This metric is defined independently of the choice of adapted
coframes. 
We notice here that the Sasaki metric was first introduced 
on the tangent bundle $TM$ \cite{sasaki-58}, but we use the 
same nomenclature on the cotangent bundle $T^*M$ as well. 

By using the Sasaki metric,
the arc length on $\TsM$ is defined as
\begin{displaymath}
  \rmd \sigma^{2} = g_{ij} \rmd x^{i} \rmd x^{j}
  + g^{ij} (\rmd p_{i} - p_{k} \Gamma^{k}_{in}\rmd x^{n})
  (\rmd p_{j} - p_{\ell} \Gamma^{\ell}_{jh}\rmd x^{h}).
\end{displaymath}
It then turns out that geodesic flows on $\TsM$ take the same arc 
length as the corresponding geodesics on $M$ have, 
since one has $\rmd\sigma^{2}=g_{ij}\rmd x^{i} \rmd x^{j}=\rmd s^{2}$ 
for horizontal curves, and since geodesic flows are horizontal. 
Hence the parameter $s$ used in Hamilton's equation
(\ref{eq:Hamilton-eq}) may be interpreted as the arc length on $M$, 
so that the geodesic  
$x(s)=\pi(P(s))$ on $M$ is described in the arc length parameter. 

We will adopt the Sasaki metric on $\TsM$ to discuss orthogonality of 
Lyapunov vectors on $\TsM$ in the next section.

%%%%%%%%%%%%%%%%%%%%%%%%%%%%%%%%%%%%%%%%%%%%%%%%
% Section 4
%%%%%%%%%%%%%%%%%%%%%%%%%%%%%%%%%%%%%%%%%%%%%%%%
\section{Lyapunov Analysis of Geodesic Flows}
\label{sec:lyapunov-analysis}

On the basis of the geometric setting-up, we are to find a first-order 
differential equation associated with the Jacobi equation, and 
thereby discuss Lyapunov vectors. 

\subsection{Linearization of Hamilton's equations of motion}
\label{sec:dynamics}

For a general Hamiltonian function $H$, 
linearized Hamilton's equations of motion 
are put, as is well-known, in the form 
\begin{displaymath}
  \begin{split}
    \dfracd{\hat{X}^{i}}{s} 
    & = \fracpp{H}{p_{i}}{x^{j}} \hat{X}^{j} 
    + \fracpp{H}{p_{i}}{p_{j}} \hat{X}^{\up{j}} , \\
    \dfracd{\hat{X}^{\up{i}}}{s} 
    & = - \fracpp{H}{x^{i}}{x^{j}} \hat{X}^{j} 
    - \fracpp{H}{x^{i}}{p_{j}} \hat{X}^{\up{j}} ,
  \end{split}
\end{displaymath}
where ${\bX}=\hat{X}^{i}\partial_{i} +
\hat{X}^{\up{i}}\partial_{\up{i}}$ stands for a deviation 
of Hamiltonian flows, where 
$\partial_{i}=\partial/\partial x^{i}$ 
and $\partial_{\up{i}}=\partial/\partial p_{i}$. 
These equations can be obtained from the condition 
$[\bX,\bX_{\!H}]=0$ as well,
where $\bX_{\!H}$ is the Hamiltonian vector field, 
\begin{displaymath}
  \bX_{\!H} = \dfracp{H}{p_{i}} \dfracp{}{x^{i}}
  - \dfracp{H}{x^{i}} \dfracp{}{p_{i}}.
\end{displaymath}
In fact, the condition $[\bX,\bX_{\!H}]=0$ restricted to a 
prescribed Hamiltonian flow $P(s)=(x(s),p(s))$ provides 
\begin{displaymath}
  \begin{split}
    [\bX,\bX_{\!H}]|_{P(s)} & = 
    \left( \fracpp{H}{p_{i}}{x^{j}} \hat{X}^{j} 
      + \fracpp{H}{p_{i}}{p_{j}} \hat{X}^{\up{j}} \right)
    \left. \dfracp{}{x^i} \right|_{P(s)} \\
    & \hspace*{2em}
    - \left( \fracpp{H}{x^{i}}{x^{j}} \hat{X}^{j} 
      + \fracpp{H}{x^{i}}{p_{j}} \hat{X}^{\up{j}} \right)
    \left. \dfracp{}{p_{i}} \right|_{P(s)} \\
    & \hspace*{2em}
    - \dfracd{\hat{X}^{i}}{s}
    \left. \dfracp{}{x^{i}} \right|_{P(s)}
    - \dfracd{\hat{X}^{\up{i}}}{s}
    \left. \dfracp{}{p^{i}}\right|_{P(s)},
  \end{split}
\end{displaymath}
where we have used the formula
\begin{displaymath}
  \dfracd{\hat{X}^{i}}{s}
  = \dfracp{H}{p_{k}} \dfracp{\hat{X}^{i}}{x^{k}}
  - \dfracp{H}{x^{k}} \dfracp{\hat{X}^{i}}{p_{k}} 
  = \bX_{\!H} (\hat{X}^{i}), 
\end{displaymath}
and a similar formula for $\rmd \hat{X}^{\up{i}}/\rmd s$. 
It is to be noted here that the condition $[\bX,\bX_{\!H}]=0$ 
implies that a Hamiltonian flow, an integral curve of $\bX_{\!H}$, 
is dragged to another infinitesimally nearby Hamiltonian flow 
by the infinitesimal transformation $\bX$, 
{\it i.e.}, $\bX$ is a deviation of Hamiltonian flows. 
With this in mind, we can obtain linearized equations with 
respect to the adapted frame, if we calculate $[\bX,\bX_{\!H}]=0$ 
with $\bX$ and $\bX_{\!H}$ expressed as 
\begin{displaymath}
  \begin{split}
    \bX & = X^{i} D_{i} + X^{\up{i}} D_{\up{i}}, \\[2mm]
    \bX_{\!H} & = D_{\up{i}}(H) D_{i} - D_{i}(H) D_{\up{i}},
  \end{split}
\end{displaymath}
respectively, and restrict the resultant equation 
to a prescribed flow $P(s)$.  
We note here that the components $(X^i,X^{\up{i}})$ with respect to 
the adapted frame transform according to 
\begin{equation}
   \label{eq:transf-comp}
    X^i=\dfracp{x^{i}}{\up{x}^{j}}\up{X}^{j}, \quad 
    X^{\up{i}}=\dfracp{\up{x}^{j}}{x^{i}}\up{X}^{\up{j}}.
\end{equation}
A long but straightforward calculation of 
$[\bX,\bX_H]|_{P(s)}=0$ then provides 
linearized Hamilton's equations of motion as follows: 
\begin{equation}
  \label{eq:deviation-general}
  \begin{split}
    \dfracd{X^{i}}{s} & = \left[
      \fracpp{H}{p_{i}}{x^{j}}
      + \fracpp{H}{p_{i}}{p_{\ell}} p_{k} \Gamma^{k}_{\ell j}
    \right] X^{j}
    + \fracpp{H}{p_{i}}{p_{j}} X^{\up{j}} , \\
    \dfracd{X^{\up{i}}}{s} & = - \left[
      \fracpp{H}{x^{i}}{x^{j}} 
      + \fracpp{H}{x^{i}}{p_{\ell}} p_{k} \Gamma^{k}_{j\ell}
      + \left(
        \fracpp{H}{p_{\ell}}{x^{j}}
        + \fracpp{H}{p_{\ell}}{p_{m}} p_{n} \Gamma^{n}_{mj}
      \right)
      p_{k} \Gamma^{k}_{i\ell}
      \right. \\
      & 
      \hspace*{3em}
      \left.
      - \dfracp{H}{x^{k}} \Gamma^{k}_{ij}
      + p_{k} \dfracp{\Gamma^{k}_{ij}}{x^{m}} \dfracp{H}{p_{m}}
    \right] X^{j}
    - \left[
      \fracpp{H}{x^{i}}{p_{j}} 
      + \fracpp{H}{p_{\ell}}{p_{j}} p_{k} \Gamma^{k}_{i\ell}
    \right] X^{\up{j}} ,
  \end{split}
\end{equation}
where use has been made of the formula 
\begin{displaymath}
  \dfracd{X^{i}}{s}
  =D_{\up{j}}(H) D_{j}(X^{i}) - D_{j}(H)D_{\up{j}}(X^{i})
  = \bX_{\!H}(X^{i}),
\end{displaymath}
and of a similar formula for $\rmd X^{\up{i}}/\rmd s$.

In what follows, we take the Hamiltonian given by Eq.(\ref{eq:Hgeo}).
The equation of deviation (\ref{eq:deviation-general}) then 
takes the form
\begin{equation}
  \label{eq:deviation-geo}
  \begin{split}
    \dfracd{X^{i}}{s}
    & = - \Gamma^{i}_{jk} g^{k \ell}p_{\ell} X^{j} + 
    g^{ij} X^{\up{j}} , \\
    \dfracd{X^{\up{i}}}{s} 
    & = - R_{jk\ell i} g^{kn}p_{n} g^{\ell h}p_{h} X^{j}
    + \Gamma^{j}_{ik} g^{k \ell}p_{\ell} X^{\up{j}} .
  \end{split}
\end{equation}
The right-hand-side of Eq.(\ref{eq:deviation-geo}) must be evaluated
along a geodesic flow $P(s)=(x(s),p(s))$.  Since one has 
$g^{ij}p_{j}(s)=\dfracd{x^i}{s}=:\xi^{i}(s)$ along the geodesic flow, 
Eq.(\ref{eq:deviation-geo}) can be brought into the form 
\begin{equation}
  \label{eq:deviation-geo2}
  \begin{split}
    \dfracd{X^{i}}{s}
    & = - \Gamma^{i}_{jk} \xi^{k} X^{j} + g^{ij} X^{\up{j}} , \\
    \dfracd{X^{\up{i}}}{s} 
    & = - R_{jk\ell i} \xi^{k} \xi^{\ell} X^{j}
    + \Gamma^{j}_{ik} \xi^{k} X^{\up{j}} .
  \end{split}
\end{equation}
We can show that this system of equations is 
the first-order differential equation 
which project to the Jacobi equation and hence may be called 
the lifted Jacobi equation.
The proof runs as follows: 
On account of (\ref{eq:transf-comp}), the quantities $(X^i(s))$ 
and $(X^{\up{i}}(s))$ may be viewed as a tangent and a cotangent vector to $M$ 
along the geodesic $x(s)$, so that the first equation of 
(\ref{eq:deviation-geo2}), rewritten as  
\begin{displaymath}
  \dfracd{X^{i}}{s} + \Gamma^{i}_{jk} \xi^{k} X^{j} 
   = g^{ij} X^{\up{j}},  
\end{displaymath} 
implies that $(g^{ij}X^{\up{j}}(s))$ is equal to the covariant derivative 
of $(X^i(s))$ along the geodesic $x(s)$. 
The second equation of (\ref{eq:deviation-geo2}) then implies that 
\begin{displaymath}
   \dfracd{Y^{i}}{s} + \Gamma^{i}_{jk} \xi^{k} Y^{j}=
   - R_{jk\ell}^{\hspace*{1.2em}i} \xi^{k} \xi^{\ell} X^{j}\quad 
   \mbox{with} \quad Y^i=g^{ij}X^{\up{j}}. 
\end{displaymath} 
The above two equations are put together to yield the Jacobi 
equation for $\bY_{\!\bX}=(X^i(s))$,
\begin{displaymath}
     \nabla_{\!\bxi}\nabla_{\!\bxi} \bY_{\!\bX} = 
     -\bR(\bY_{\!\bX},\bxi)\bxi,
\end{displaymath}
where $\nabla_{\bxi}$ stands for the covariant derivation along 
the geodesic $x(s)$. 

%%%%%%%%%%%%%%%%%%%%%%%%
% Subsection IV B
%%%%%%%%%%%%%%%%%%%%%%%%
\subsection{Lyapunov vectors}
\label{sec:lyapunov-vectors}

Here we show that solutions to Eq.(\ref{eq:deviation-geo2}) satisfy
the requirement stated in Introduction in the Hamiltonian system with
the Hamiltonian $K$ given in (\ref{eq:Hgeo}).
As for the gradient of $K$, we note that the differential $\rmd K$ and 
the gradient of $K$, $\gradK$, are put in the form
\begin{displaymath}
   \rmd K = g^{ik} p_{k} \theta^{\up{i}}, \qquad 
   \mbox{grad}K = p_{i} D_{\up{i}}, 
\end{displaymath}
respectively, where the gradient of a function $F$ on $\TsM$,
$\text{grad}F$, is defined through
\begin{displaymath}
  \til{\bg}(\text{grad}F, \bX) = \rmd F(\bX) 
\end{displaymath}
for any vector field $\bX\in\Xbar(\TsM)$.

It is an easy matter to verify that Eq.(\ref{eq:deviation-geo2}) is
satisfied by $\bX_{K}$, the tangent vector to a Hamiltonian flow
$P(s)$ or a geodesic flow in $\TsM$. In fact, the tangent vector
$\bX_{K}$ to $P(s)$ is given (\ref{eq:geod-spray}),  
and has the components,
$X^{i}(s)=g^{i\ell}p_{\ell}(s),\,X^{\up{i}}(s)=0$, 
satisfying (\ref{eq:deviation-geo2}).
While the gradient vector along the Hamiltonian flow $P(s)$,
which is denoted by $\mbox{grad}K(s)$ for simplicity, 
is not a solution to the linearized equation
(\ref{eq:deviation-geo2}), the vector
$\mbox{grad}K(s) + s \bX_{\!K}(s) 
= p_{i}(s) D_{\up{i}} + s\ g^{ik} p_{k}(s) D_{i}$
is a solution to (\ref{eq:deviation-geo2}), as is easily verified. 
Taking this into account, we wish to decompose the tangent space 
$T_{P(s)}\TsM$ to $\TsM$ at every point $P(s)$ of a geodesic flow 
into the direct sum of the plane spanned by both $\bX_{\!K}(s)$ 
and $\mbox{grad}K(s)$ and the subspace transversal to the plane. 
Let us define subspaces $N_{P(s)}$ and $E_{P(s)}$ to be 
\begin{equation}
  \label{eq:N&E}
  \begin{split}
    N_{P(s)} 
    & = \{ {\bX} \in T_{P(s)}\TsM
    \ |\ 
    {\bX} = \alpha\, \bX_{\!K}(s) + \beta\, \mbox{grad}K(s),
    \; \alpha,\, \beta \in\bR \} ,\\[2mm]
    E_{P(s)}
    & = \{ {\bX} \in T_{P(s)}\TsM
    \ |\ 
    \til{\bg}({\bX},\bX_{\!K}(s)) =0, 
    \; \til{\bg}(\bX,\mbox{grad}K(s)) =0 \} ,
  \end{split}
\end{equation}
respectively, where $E_{P(s)}$ is the orthogonal complement 
of $N_{P(s)}$ with respect to the Sasaki metric $\til{\bg}$. 
Thus we have the orthogonal direct sum decomposition, 
\begin{equation}
   \label{eq:decomp-L}
  T_{P(s)}\TsM=N_{p(s)} \oplus E_{P(s)}.
\end{equation}
We wish to show that these subspaces are invariant under any solution to 
the linearized equation (\ref{eq:deviation-geo2}).  
To this end, we have to verify: 
\medskip
\par\noindent
{\bf Theorem}: 
A solution $\bX(s)$ to the linearized equation (\ref{eq:deviation-geo2})
which is in $N_{P(0)}$ (resp. in $E_{P(0)}$) at an initial moment $s=0$ 
keeps belonging to $N_{P(s)}$ (resp. to $E_{P(s)}$) at any instant $s$.  

\medskip
\par\noindent
The proof of this statement is carried out as follows: 
As we have already shown, $\bX_{\!K}(s)$ and 
$\mbox{grad}K(s)+s\,\bX_{\!K}(s)$ are solutions to 
(\ref{eq:deviation-geo2}), so that the linear combination of them, 
$\alpha\, \bX_{\!K}(s)+\beta(\mbox{grad}K(s)+s\,\bX_{\!K}(s))=
(\alpha+\beta s)\bX_{\!K}(s)+\beta\,\mbox{grad}K(s)$, is also 
in $N_{P(s)}$ at any instant $s$, which proves the invariance of
$N_{P(s)}$ under the linearized flow $\bX(s)$. 
To prove the invariance of $E_{P(s)}$, we consider the temporal 
evolution of $\til{\bg}({\bX},\bX_{\!K})$ with ${\bX}$ a solution to 
(\ref{eq:deviation-geo2}). 
We are to show that 
\begin{equation}
 \label{eq:temp-evol}
   \begin{split}
   \dfracd{}{s}\til{\bg}({\bX}, \bX_{\!K})
   & = \rmd K (\bX),\\
   \frac{\rmd^2}{\rmd s^2}\til{\bg}({\bX},\bX_{\!K})
   & = 0.
   \end{split}
\end{equation}
We can carry out the proof of these equations in the manner 
of mechanics as follows: 
Note that $\til{\bg}(\bX,\bX_{\!K})=\theta(\bX)$, 
where $\theta$ is the standard one-form on $\TsM$,
{\it i.e.}, $\theta=p_{i} \rmd x^{i}$ in local coordinates. 
Then differentiation of $\theta(\bX)$ with respect to $s$ results in 
\begin{displaymath}
  \begin{split}
    \dfracd{}{s} \theta(\bX) 
    & = {\cal L}_{\bX_{\!K}}(\theta(\bX)) \\[2mm]
    & = ({\cal L}_{\bX_{\!K}}\theta)(\bX) + \theta([\bX_{\!K},\bX])\\[2mm]
    & =(\rmd \iota(\bX_{\!K}) \theta +
    \iota(\bX_{\!K}) \rmd\theta)(\bX)\\[2mm]
    & = (\rmd(\theta(\bX_{\!K})) - \rmd K)(\bX)\\[2mm]
    & =\rmd K(\bX),
   \end{split}
\end{displaymath}
where use has been made of (i) the definition of the Lie derivative 
of one-forms, (ii) the condition $[\bX,\bX_{\!K}]=0$, 
(iii) the Cartan's formula for the Lie derivation, 
(iv) $\iota(\bX_{\!K})\rmd\theta =-\rmd K$, and 
(v) the equality $\theta(\bX_{\!K})=2K$ due to the homogeneity 
of $K$ in $p_{i}$. 
Thus we obtain the former equation of (\ref{eq:temp-evol}). 
Differentiating the former equation of (\ref{eq:temp-evol})
with respect to $s$ using the equation
\begin{displaymath}
  \dfracd{}{s} \rmd K(\bX) =
  \dfracd{}{s} \til{\bg}(\gradK,\bX) = 0,
\end{displaymath}
a similar equation to (\ref{eq:inpro-X-gradH}),
we obtain the latter of (\ref{eq:temp-evol}). 
Now Eq.(\ref{eq:temp-evol}) is integrated to give 
\begin{displaymath}
  \til{\bg}(\bX,\bX_{\!K})|_{P(s)}
  = \til{\bg}(\bX,\bX_{\!K})|_{P(0)}
  + s\, \rmd K (\bX)|_{P(0)}. 
\end{displaymath}
Since $\rmd K(\bX)=\til{\bg}(\bX,\mbox{grad}K)$,
the above equation implies that $\bX(s)\in E_{P(s)}$ 
if $\bX(0)\in E_{P(0)}$. 
This ends the proof of the invariance of $E_{P(s)}$ 
under the linearized flow $\bX(s)$. 
$\blacksquare$

\vspace*{1em}

On the basis of the decomposition (\ref{eq:decomp-L}), 
we can construct a set of 
Lyapunov vectors $\{\bV_{a}\},\,a=1,\cdots,2m$, 
satisfying the requirement mentioned in Sec.\ref{sec:introduction}. 
We are thinking of the Riemannian manifold 
$(M_{J},\bg_{J})$ introduced in Sec.\ref{sec:jacobi-metric},
and hence $m=N$. 
The first $N-1$ linearly independent solutions, $\{\bX_a(s)\},\,
a=1,\cdots,N-1$, to the lifted Jacobi equation (\ref{eq:deviation-geo2}) 
are chosen in $E_{P(s)}$, which are orthogonalized to give first 
$N-1$ Lyapunov vectors $\{\bV_{1},\cdots,\bV_{N-1}\}$. 
The $N$-th and $(N+1)$-th Lyapunov vectors are chosen in $N_{P(s)}$ 
so as to be $\bV_N(s)=\bX_{\!K}(s)$ and $\bV_{N+1}(s)=\mbox{grad}K(s)$, 
respectively.
This is because they are mutual orthogonal and because $\bX_{\!K}(s)$
and $\mbox{grad}K(s)+s\ \bX_{\!K}(s)$ are solutions to the linearized  
equation and further orthogonal to the first $N-1$ Lyapunov vectors.
Note in addition that any solution $\bX(s)$ staying in $N_{P(s)}$ 
becomes asymptotically parallel to $\bX_{\!K}(s)$ as $s \to \infty$, 
so that $\bX_{\!K}(s)$ is assigned to the $N$-th Lyapunov vector 
and $\gradH(s)$ to the $(N+1)$-th 
Lyapunov vector, respectively. 
The remaining $N-1$ Lyapunov vectors are chosen in $E_{P(s)}$, 
which are orthogonal to $\bX_{\!K}$ and $\mbox{grad}K$ 
as well as to the first $N-1$ Lyapunov vectors by the very definition. 
Consequently, from solutions to (\ref{eq:deviation-geo2}) 
with the initial values chosen so as to satisfy 
\begin{enumerate}
\item[(a)]
$\bX_N(0)=\bX_{\!K}(0),\quad \bX_{N+1}(0)=\mbox{grad}K(0)$,
\item[(b)]
$\bX_{a}(0)\perp \{\bX_{\!K}(s),\mbox{grad}K(0)\},\quad 
a=1,\cdots,N-1,N+2,\cdots,2N$,
\end{enumerate}
at the initial moment $s=0$, 
we can obtain expectedly a set of Lyapunov vectors 
such that 
\begin{enumerate}
\item[(i)]
$\bV_N(s)=\bX_{\!K}(s),\quad \bV_{N+1}(s)=\mbox{grad}K(s)$,
\item[(ii)]
$\bV_{a}(s)\perp \{\bX_{\!K}(s),\mbox{grad}K(s)\},\quad 
a=1,\cdots,N-1,N+2,\cdots,2N$.
\end{enumerate}

From the property (i), we can observe that the Lyapunov exponents 
$\lambda_N$ and $\lambda_{N+1}$ vanish indeed. In fact, since 
\begin{displaymath}
   \til{\bg}(\bX_{\!K},\bX_{\!K})=\til{\bg}(\gradK,\gradK)=2K
\end{displaymath}
is constant along any geodesic flow, one has $\lambda_{N}=\lambda_{N+1}=0$ 
from the formula (\ref{eq:Lyapunov-exponents}). 

%%%%%%%%%%%%%%%%%%%%%%%%%%%%%%%%%%%%%%%%%%%%%%%%
% Section 5
%%%%%%%%%%%%%%%%%%%%%%%%%%%%%%%%%%%%%%%%%%%%%%%%
\section{Numerical Calculations for Comparison}
\label{sec:calculation}

In this section, 
we are to compare the geometric method and the usual method 
through a model system with $3$ degrees of freedom, 
by numerically calculating Lyapunov exponents and Lyapunov vectors 
in respective methods. 
We will find that the Lyapunov exponents calculated in respective 
methods coincide with each other, independently of 
the choice of methods, while the Lyapunov vectors calculated on 
respective setting-ups exhibit different behaviors to each other, 
depending on the method chosen.

\subsection{Comparison of setting-ups in respective methods}
\label{sec:geo-and-usual}

\begin{table}[htbp]
  \begin{center}
    \caption{Comparison between the usual method and the geometric
      method. The $N$-dimensional manifold $M_{J}$ is defined
      in Eq.(\protect\ref{eq:Jacobi-mfd}),
      and $\til{\bg}_{E}$ is the Euclidean metric.
      Note that $\TsM_{J}=M_{J}\times\bR^{N}$.}
    \begin{tabular}{cccccc}
      Method & Configuration & Phase & Metric & Hamiltonian &
      Linearized\\ 
      & space & space & & & equation \\
      \hline
      Usual & $M_{J}$ & $M_{J}\times\bR^{N}$ & $\til{\bg}_{E}$
      & $H$ [Eq.(\ref{eq:natural-Hamiltonian})] &
      Eq.(\ref{eq:linearized-Hamilton-eq}) \\
      Geometric & $M_{J}$ & $\TsM_{J}$ & $\til{\bg}_{J}$
      & $K$ [Eq.(\ref{eq:Hgeo})] &
      Eq.(\ref{eq:deviation-geo2}) 
    \end{tabular}
    \label{tab:geo-and-usual}
  \end{center}
\end{table}

For a natural Hamiltonian system with $N$ degrees of freedom, 
setting-ups for Lyapunov analysis both in the geometric method and in 
the usual method are summarized in Tab.\ref{tab:geo-and-usual}.
We note here that the metric $\til{\bg}_E$ introduced on the phase space 
$M_J\times \bR^N$ in the usual method is, of course, the Euclidean metric 
defined, as usual, to be 
\begin{displaymath}
  \til{\bg}_E = \delta_{ij} \rmd q^{i} \otimes \rmd q^{j}
  + \delta^{ij} \rmd p_{i} \otimes \rmd p_{j}.
\end{displaymath}

As was pointed out in Sec.\ref{sec:jacobi-metric}, 
the geodesic equations for the Jacobi metric are equivalent to 
Newton's equations of motion for a natural dynamical system with
energy $E$.
We now verify this fact in the Hamiltonian formalism. 
The Hamiltonian vector fields $\bX_{\!K}$ and $\bX_{\!H}$ which 
are defined on the same phase space in respective manners are given by 
\begin{equation}
  \label{eq:JE-Ham-vects1}
  \bX_{\!K} = g^{ij} p_{j} D_{i}, 
    \quad
  \bX_{\!H} = \delta ^{ij}\Bigl( p_{j} \dfracp{}{q^{i}}
  - \dfracp{V}{q^{j}} \dfracp{}{p_{i}}\Bigr),
\end{equation}
respectively, 
where $g^{ij}=\delta^{ij}/2(E-V)$. 
A straightforward calculation along with (\ref{eq:Gamma-J}) 
and $\frac12\sum p_i^2+V=E$ then provides 
\begin{equation}
  \label{eq:JE-Ham-vects2}
  \bX_{\!K}= \frac{1}{2(E-V)}\bX_{\!H},
\end{equation}
which implies that Hamiltonian flows both in the geometric method 
and in the usual method coincide within the change of parameters, 
$\rmd s/\rmd t=2(E-V(q))$. 
Thus, along the same flow (up to the parameter change), 
we can compare numerically tangent vectors such as solutions to 
linearized equations of motion and Lyapunov vectors. 
In the following, $\bX^{(g)}(s)$ and $\bX(t)$ denote solutions to 
the linearized equations of motion in the geometric method 
and in the usual method, respectively.

\subsection{Orthogonal relations in the usual method}
\label{sec:orthgonality-usual}

In Sec.\ref{sec:lyapunov-analysis}, 
we have shown that Lyapunov vectors in the geometric method 
can be chosen so that two of them may be the tangent vectors 
to the Hamiltonian flow in question and the gradient vector of 
the Hamiltonian function along the flow, and the others be orthogonal
to those two vectors. 
In this subsection, we remark that such orthogonal relations 
holds for part of Lyapunov vectors even in the usual method, 
in which the Euclidean metric $\til{\bg}_E$ is adopted 
in $M_{J}\times\bR^{N}$. 

Let $\bX_{1}(t),\cdots,\bX_{2N}(t)$ be linearly independent 
solutions to Eq.(\ref{eq:linearized-Hamilton-eq}), for which 
the initial conditions are taken in such a manner that
\begin{enumerate} 
\item[(a)]
 $\bX_{N}(0)=\bX_{\!H}(0),\quad \bX_{N+1}(0)=\gradH(0)$,
\item[(b)] 
 $\bX_{a}(0) \perp \{ \bX_{\!H}(0),\; \gradH(0)\},
  \quad a=1,\cdots,N-1,N+2,\cdots,2N,$ 
\end{enumerate}
where $\bX_{\!H}$ and $\gradH$ are the Hamiltonian vector 
field for $H$ and the gradient vector field of $H$, respectively. 
Let $\bV_{1}(t),\cdots,\bV_{2N}(t)$ 
be Lyapunov vectors formed from $\bX_{a}(t),\,a=1,\cdots,2N$. 
Then the following two properties hold true: 
\begin{enumerate}
\item[(i)]
 $\bV_{1}(t),\cdots,\bV_{N}(t)$ are always orthogonal to 
 $\gradH(t)$, 
\item[(ii)]
 $\bV_{N+1}(t),\cdots,\bV_{2N}(t)$ are always orthogonal to
 $\bX_{\!H}(t)$.
\end{enumerate}

The property (i) is easily shown to hold from
Eq.(\ref{eq:inpro-X-gradH}). 
In fact, solutions $\bX_{1}(t),\cdots,\bX_{N}(t)$ to the linearized 
equations (\ref{eq:linearized-Hamilton-eq}) are always orthogonal to
$\gradH(t)$, if they are initially orthogonal to $\mbox{grad}H(0)$. 
Hence the Lyapunov vectors $\bV_{1}(t),\cdots,\bV_{N}(t)$
are always orthogonal to $\gradH(t)$, 
since the $N$-dimensional space spanned by
$\bV_{1}(t),\cdots,\bV_{N}(t)$
is the same as that spanned by $\bX_{1}(t),\cdots,\bX_{N}(t)$. 
For the proof of the property (ii),
we use the fact that the Hamiltonian vector field $\bX_{\!H}(t)$ is a
solution to the linearized Hamilton's equation
(\ref{eq:linearized-Hamilton-eq}), so that one has  
$\bX_{N}(t)=\bX_{\!H}(t)$. 
Then $\bX_{\!H}(t)$ is in the $N$-dimensional space spanned by
$\bX_{1}(t),\cdots,\bX_{N}(t)$, 
and hence in that spanned by $\bV_{1}(t),\cdots,\bV_{N}(t)$.
By definition, 
the Lyapunov vectors $\bV_{N+1}(t),\cdots,\bV_{2N}(t)$ are 
orthogonal to $\bV_{1}(t),\cdots,\bV_{N}(t)$, 
and hence to $\bX_{\!H}(t)$. 

The above two properties will be confirmed as well by numerical 
calculations for a model system in a later subsection.
Moreover, by numerical calculations in the usual method, 
we will observe that 
$\bV_{N+2}(t),\cdots,\bV_{2N}(t)$ are not always orthogonal to
$\gradH(t)$, and that $\bV_{1}(t),\cdots,\bV_{N-1}(t)$ 
are not always orthogonal to $\bX_{\!H}(t)$, either. 
We recall here that, in the geometric method, Lyapunov 
vectors $\bV^{(g)}_{N+2}(s),\cdots,\bV^{(g)}_{2N}(s)$ are 
always orthogonal to $\text{grad}K(s)$, and that 
$\bV^{(g)}_{1}(s),\cdots,\bV^{(g)}_{N-1}(s)$ are always 
orthogonal to $\bX_{\!K}(s)$, which will be confirmed 
as well by numerical calculations for the model system.
Here, by $\bV_{a}^{(g)}$ and $\bV_{a}$, we denote the Lyapunov
vectors which are obtained in the geometric method and in the usual
method, respectively, to tell the difference between them.

\subsection{Initial conditions} 
\label{sec:initial-conditions}

To compare numerical computation results calculated both 
in the geometric method and in the usual method, we have to  
set both Hamilton's equations of motion and linearized equations 
of motion to share the same initial conditions. 
Hence, in particular, we come to require that the initial conditions 
for linearized equations of motion are taken to be subject to the 
conditions (a) and (b) mentioned in Sec.\ref{sec:orthgonality-usual} 
in the usual method as well as in the geometric method. 
In this subsection, we discuss how one can set such initial
conditions, in spite of the difference between metrics used. 

We take a number of initial values, $P(0)=(q^{j}(0),p_{j}(0))$, 
for Hamiltonian flows on $\TsM_{J}$ in such a manner
that $\text{grad}V$ vanishes at the initial point $P(0)$, 
where $\text{grad}V$ is defined with respect to both the Euclidean 
metric and the Jacobi metric on the configuration space $M_J$, 
but the equation $\text{grad}V=0$ defines the same points, 
independently of the metric chosen. 
Since the phase spaces in both methods are in common, and since 
Hamiltonian flows in both methods are also in common up to 
the change of parameters, we will obtain a number of Hamiltonian 
flows in common after integration. 
We have also to note that the condition $\text{grad}V=0$ at the 
initial point implies that the Christoffel symbols $\Gamma^i_{jk}$'s 
defined by (\ref{eq:Gamma-J}) vanish also there, 
so that the Jacobi metric is put, at the initial point, in the form 
\begin{equation}
   \label{eq:at-initial-pt}
   \til{\bg}_J|_{P(0)}=2W_0\delta_{ij}\rmd x^i\otimes\rmd x^j
             +(2W_0)^{-1}\delta^{ij}\rmd p_i\otimes\rmd p_j,
\end{equation}
where $W_0=E-V(P(0))$. 

Let $\bX_{a}^{(g)}$ and $\bX_{a}$ denote solutions to linearized
equations (\ref{eq:deviation-geo2}) and
(\ref{eq:linearized-Hamilton-eq}), respectively.

According to the procedure stated in Sec.\ref{sec:lyapunov-analysis}, 
initial conditions for linearized equations (\ref{eq:deviation-geo2}) 
in the geometric method are set as follows: 
\begin{enumerate}
\item[(a)]
  $\bX^{(g)}_{N}(0)=\bX_{\!K}(0),\quad 
  \bX^{(g)}_{N+1}(0)=\text{grad}K(0)$, 
\item[(b)]
  $\bX^{(g)}_{a}(0) \in E_{P(0)} \cap H_{P(0)}, \quad 
  a=1,\cdots, N-1,\\ 
   \bX^{(g)}_{b}(0) \in E_{P(0)} \cap V_{P(0)}, \quad b=N+2,\cdots, 2N$. 
  \end{enumerate}
See (\ref{eq:decomp}) and (\ref{eq:N&E}) for the definitions of 
$H_{P(s)}$, $V_{P(s)}$, and $E_{P(s)}$.

Initial conditions for the linearized equations
(\ref{eq:linearized-Hamilton-eq})
in the usual method are set as 
\begin{displaymath}
   \bX_{a}(0)=\bX^{(g)}_{a}(0),\quad a=1,\cdots,2N.
\end{displaymath}
We here have to verify that these initial vectors
$\bX_{a}(0),\,a=1,\cdots,2N$, 
are indeed subject to the initial conditions (a) and (b)
stated in Sec.\ref{sec:orthgonality-usual}. 
The verification of this is carried out as follows: 
By definition, one has $\bX_{N}(0)=\bX_{\!K}(0)$, and further 
$\bX_{\!K}(0)=\bX_{\!H}(0)/2(E-V(q(0)))$ from (\ref{eq:JE-Ham-vects2}), 
so that $\bX_{N}(0)=\bX_{\!H}(0)/2(E-V(q(0)))$. 
The constant factor $2(E-V(q(0)))$ causes no serious problem, 
since we are interested in orthogonal relations between initial vectors. 
Moreover, it is an easy matter to see that
$\bX_{N+1}(0)=\text{grad}K(0)=\text{grad}H(0)$ 
on account of the assumption $\text{grad}V(0)=0$ at the initial point, 
where we note that $\text{grad}K$ and $\text{grad}H$ are taken 
with respect to metrics, $\til{\bg}_{J}$ and $\til{\bg}_E$, respectively. 
To verify that the other initial vectors,
$\bX_{a}(0),\,a=1,\cdots,N-1,N+2,\cdots,2N$, 
are orthogonal to $\bX_{\!H}(0)$ and to $\text{grad}H(0)$, 
we use the following four facts:
(i) $\bX_{1}(0),\cdots,\bX_{N-1}(0)\in E_{P(0)}\cap H_{P(0)}$,
(ii) $\bX_{N+2}(0),\cdots,\bX_{2N}(0)\in E_{P(0)}\cap V_{P(0)}$,
(iii) $H_{P(0)}$ and $V_{P(0)}$ are orthogonal with respect to 
the Euclidean metric, as is seen from (\ref{eq:at-initial-pt}), 
(iv) restricted to the subspaces $H_{P(0)}$ and $V_{P(0)}$, 
the Jacobi metric and the Euclidean metric are conformal to each other;  
\begin{displaymath}
  \begin{split}
  \til{\bg}_J|_{P(0)}(\bX_{1},\bX_{2})
  & = 2(E-V) \til{\bg}_E|_{P(0)}(\bX_{1},\bX_{2}),
  \quad \bX_{1},\bX_{2}\in H_{P(0)},\\[2mm]
  \til{\bg}_J|_{P(0)}(\bX_{1},\bX_{2}) & =
  \til{\bg}_E|_{P(0)}(\bX_{1},\bX_{2})/2(E-V),
  \quad \bX_{1},\bX_{2}\in V_{P(0)}.
  \end{split}
\end{displaymath}
It then turns out from (i) and (iv) that $\bX_1(0),\cdots,\bX_{N-1}(0)$ 
are also orthogonal to $\bX_{N}(0)$ with respect to $\til{\bg}_E|_{P(0)}$, 
and further from (ii) and (iii) that they are also orthogonal to 
$\bX_{N+1}(0)$ with respect to $\til{\bg}_E|_{P(0)}$ .  
A similar statement for $\bX_{N+2}(0),\cdots,\bX_{2N}(0)$ holds true.
%from (ii), (iii), and (iv). 

\subsection{A model system} 
\label{sec:model}

The model system we are to consider here is a natural Hamiltonian system 
with $3$ degrees of freedom 
which has interactions of H\'enon-Heiles type,
\begin{equation}
  \label{eq:model}
  \begin{split}
    & H(q,p) = \sum_{i=1}^{3} \left[ 
      \frac{1}{2} p_{i}^{2} + V_{HH}(q^{i},q^{i+1})
    \right] , \\
    & V_{HH}(x,y) = x^{2} y - \dfrac{1}{3} y^{3},
  \end{split}
\end{equation}
where $q^{4}=q^{1}$. 
Hamiltonian vector fields both in the geometric method 
and in the usual method, denoted by $\bX_{\!K}$ and $\bX_{\!H}$, 
respectively, are given by (\ref{eq:JE-Ham-vects1}) 
with $g^{ij}=\delta^{ij}/2(E-V)$ and $V=\sum_{i=1}^{3}
V_{HH}(q^{i},q^{i+1})$.

Hamiltonian flows of $\bX_{\!H}$ for the Hamiltonian (\ref{eq:model}) 
are numerically calculated by the use of the $4$-th order symplectic 
integrator \cite{yoshida-90}, which is a numerical integration method 
on the basis of discrete time evolution with each step an explicit 
symplectic mapping.  
Initial conditions for Hamilton's equations of motion are set as 
$q^{i}(0)=0$ and $p_{i}(0)=\alpha\gamma_{i} \ (j=1,2,3)$, 
where $\gamma_{i}$'s are random values obtained from 
the uniform distribution function on the interval $[0,1]$, 
and the constant $\alpha$ is determined so as to satisfy 
the energy condition $\sum_{i=1}^{3} (p_{i}(0))^{2}/2=E$. 
For the initial values $q^{i}(0)=0$, 
we verify easily that the condition $\text{grad}V(0)=0$ is satisfied, 
which was assumed in the previous subsection \ref{sec:initial-conditions}. 
To integrate the linearized Hamilton's equations of motion 
(\ref{eq:linearized-Hamilton-eq}), we take an alternative method, that is, 
we choose to linearize, along a certain Hamiltonian flow, the sequence of 
symplectic mappings already obtained on the symplectic integrator algorithm. 
To our knowledge of explicit symplectic integrators,
the symplectic integrator used here and another symplectic algorithm 
proposed in \cite{casetti-95b} are set up on the assumption that 
Hamiltonians are of the form $H(q,p)=T(p)+V(q)$, 
so that those algorithms are not applicable to 
the numerical integration of Hamilton's equations of motion with Hamiltonians 
of the form $K(q,p)=\sum_{i=1}^{3}p_{i}^{2}/4(E-V(q))$.  
This means that we have to take another algorithm to integrate Hamilton's 
equations of motion in our geometric method for Lyapunov analysis. 
What we use in this article is an implicit but symplectic $6$th-order 
Runge-Kutta method (Kuntzmann \& Butcher method \cite{hairer-93}, 
see Appendix \ref{sec:appendixB}). 
However, we have to note here that we do not need to apply that Runge-Kutta 
method to integrate numerically Hamilton's equations of motion for $K$, 
since the solutions to Eq.(\ref{eq:Hamilton-eq}) coincide with 
Hamiltonian flows already obtained by the explicit symplectic integrator 
up to the parameter change. 
We apply the implicit Runge-Kutta method to the numerical integration of 
the lifted Jacobi equations (\ref{eq:deviation-geo2}), the linearized 
Hamilton's equations of motion for $K$. 
The implicit Runge-Kutta method, however, requires an additional process 
of numerical computation.  
In fact, we need to calculate the inverse of a $6N\times 6N$ matrix 
at each step of the integration, where $N$ denotes the 
degrees of freedom.  For this reason, the CPU-time we have needed 
to integrate the lifted Jacobi equations 
by the implicit Runge-Kutta algorithm is about $26$ times as long as 
the CPU-time we have needed to integrate the linearized Hamilton's equations
of motion for $H$ by the explicit symplectic integrator. 
We have set the unit time slice as wide as $h=2.5\times 10^{-6}$
both for the explicit symplectic integrator and the Runge-Kutta
algorithm.

\subsection{Results of numerical calculations}
\label{sec:results}

\begin{figure}[htbp]
  \begin{center}
    \includegraphics[width=10.0cm,clip,keepaspectratio]{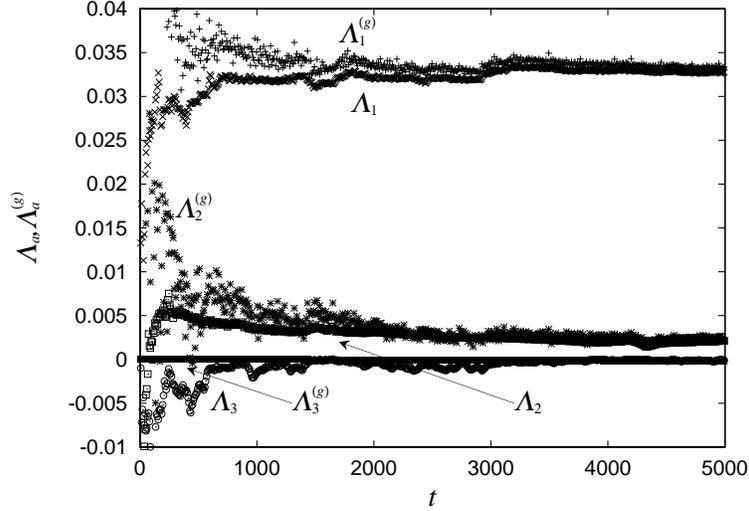}
    \caption{Convergence of Lyapunov exponents with $E=0.04$.
      Curves represent graphs of $\Lambda_{a}^{(g)}$ and $\Lambda_{a}$
      ($a=1,2,3$),
      where $\Lambda_{a}^{(g)}$ and $\Lambda_{a}$, functions in the 
      time parameter, are obtained by the geometric method and 
      by the usual method, respectively.}
    \label{fig:convergence}
  \end{center}
\end{figure}

\begin{figure}[htbp]
  \begin{center}
    \includegraphics[width=10.0cm,clip,keepaspectratio]{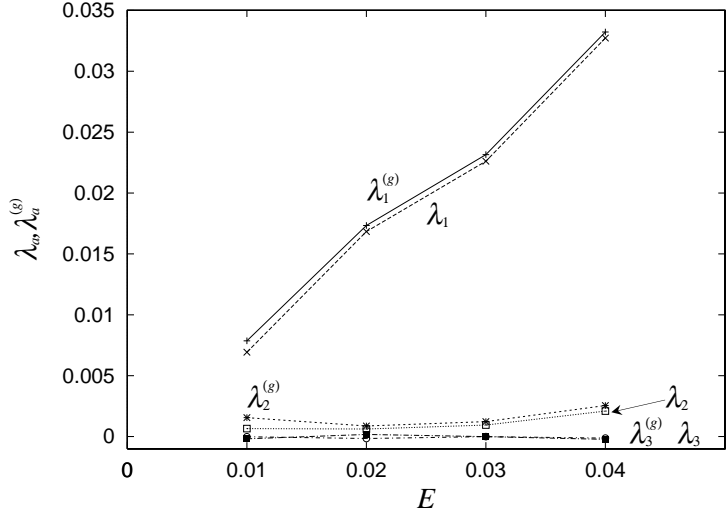}
    \caption{Comparison of Lyapunov exponents obtained
      by both the geometric method and the usual method.
      By $\lambda_{a}^{(g)}$ and $\lambda_{a}$, we denote 
      Lyapunov exponents obtained by the geometric method and 
      by the usual method, respectively.  Numerical results obtained in 
      both methods are in good agreement.}
    \label{fig:LSgeodyn}
  \end{center}
\end{figure}

Figure \ref{fig:convergence} shows that Lyapunov exponents 
calculated in both methods 
have indeed definite values for $E=0.04$,
where $\Lambda_{a}$'s and $\Lambda^{(g)}_{a}$'s are defined, 
respectively, to be 
\begin{displaymath}
  \Lambda^{(g)}_{a}(t) = \dfrac{1}{t} 
  \ln \dfrac{ \norm{V_{a}(s(t))} }{ \norm{V_{a}(0)} },
  \quad 
  \Lambda_{a}(t) = \dfrac{1}{t} 
  \ln \dfrac{ \norm{V_{a}(t)} }{ \norm{V_{a}(0)} },
   \quad a=1,2,3,
\end{displaymath}
which are supposed to be convergent to Lyapunov exponents;  
$\lim_{t\to\infty} \Lambda_{a}(t) = \lambda_{a}$ and 
$\lim_{t\to\infty} \Lambda^{(g)}_{a}(t) = \lambda^{(g)}_{a}$. 
Here, the quantities with the superscript $(g)$ are those used 
in the geometric method. 
However, to compare the numerical results, we have made $\Lambda^{(g)}_{a}(s)$ 
into a function of $t$ by means of the parameter change. 
It is to be noted here that $\Lambda_{3}^{(g)}(s)$ always vanishes 
on account of the fact that $\norm{\bX^{(g)}_{3}}=\norm{\bX_{K}}=2K$=constant.
For $E=0.01,\ 0.02$ and $0.03$, we have obtained also definite 
Lyapunov exponents, which are shown in Fig.\ref{fig:LSgeodyn} 
along with the dependence on energy. 
Figure \ref{fig:LSgeodyn} also shows that the Lyapunov exponents, 
$\lambda_a$ and $\lambda^{(g)}_a$, calculated in both methods coincide 
with each other, which means that the Lyapunov exponents are obtained 
independently of the choice of methods, geometric or usual.

We remark here that if one uses the Jacobi equations (\ref{eq:Jacobi-eq}), 
a second-order differential equation, to calculate the exponential growth 
rates of trajectories, one may obtain the same value 
as that obtained in the usual method.  
For example, for the Fermi-Pasta-Ulam $\beta$ model, 
the largest Lyapunov exponent is calculated by using a $2N$-dimensional 
vector $(X^{i},\rmd X^{i}/\rmd t)$ \cite{cerruti-sola-97}, 
where $(X^{i}(t))$ is a solution to the Jacobi equations 
(\ref{eq:Jacobi-eq}) and the Euclidean metric is 
used for the $2N$-dimensional vector.  According to \cite{cerruti-sola-97}, 
the resultant value of the exponent coincides with the largest Lyapunov 
exponent obtained in the usual method. 
This might suggest that to calculate the largest Lyapunov exponent, 
one does not need to work with the cotangent spaces. 
However, the advantage of the geometric method developed in this article 
is that after the geometric method, we can obtain all the Lyapunov exponents 
along with the Lyapunov vectors among which two vectors associated with 
the vanishing Lyapunov exponents can be separated out from the others. 
This can be observed in Figs.3 and 4. 

\begin{figure}[htbp]
  \begin{center}
    \subfigure[$1$st vector]{
      \includegraphics[width=5.5cm,clip,keepaspectratio]{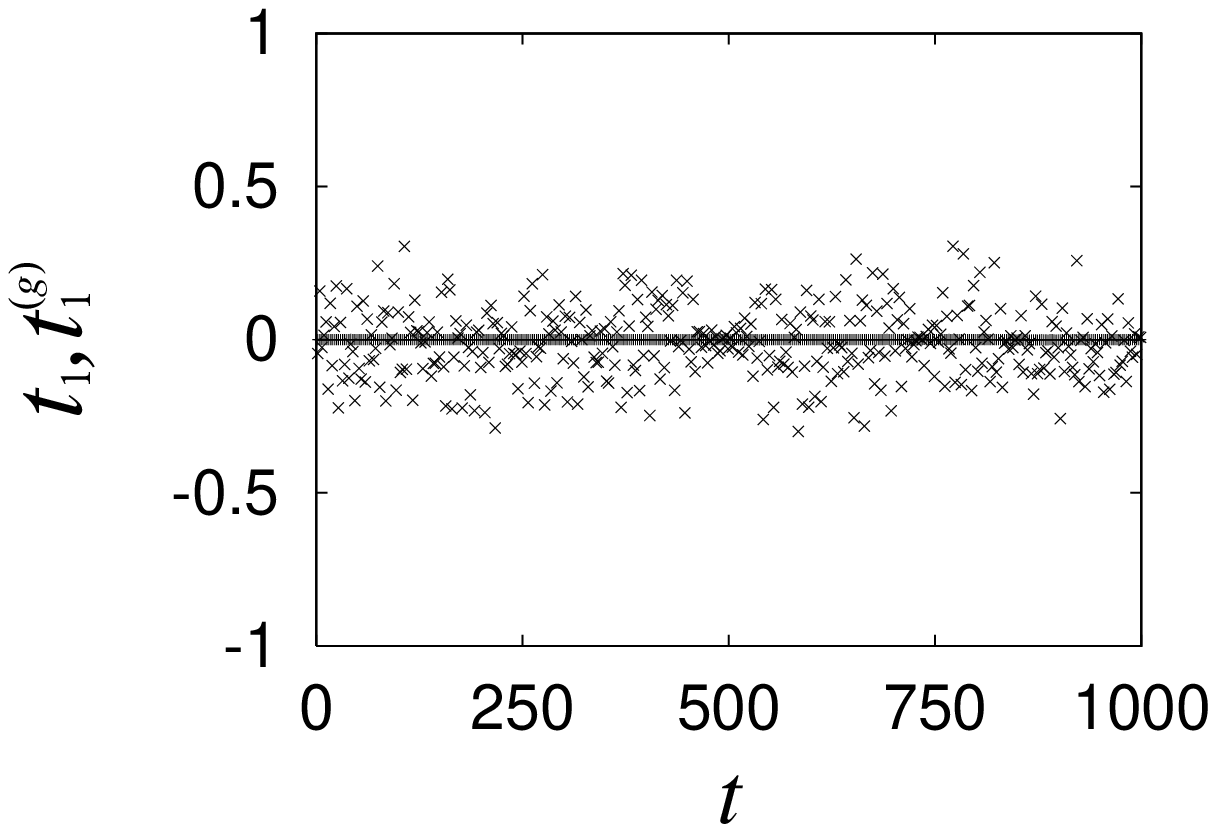}}
    \subfigure[$2$nd vector]{
      \includegraphics[width=5.5cm,clip,keepaspectratio]{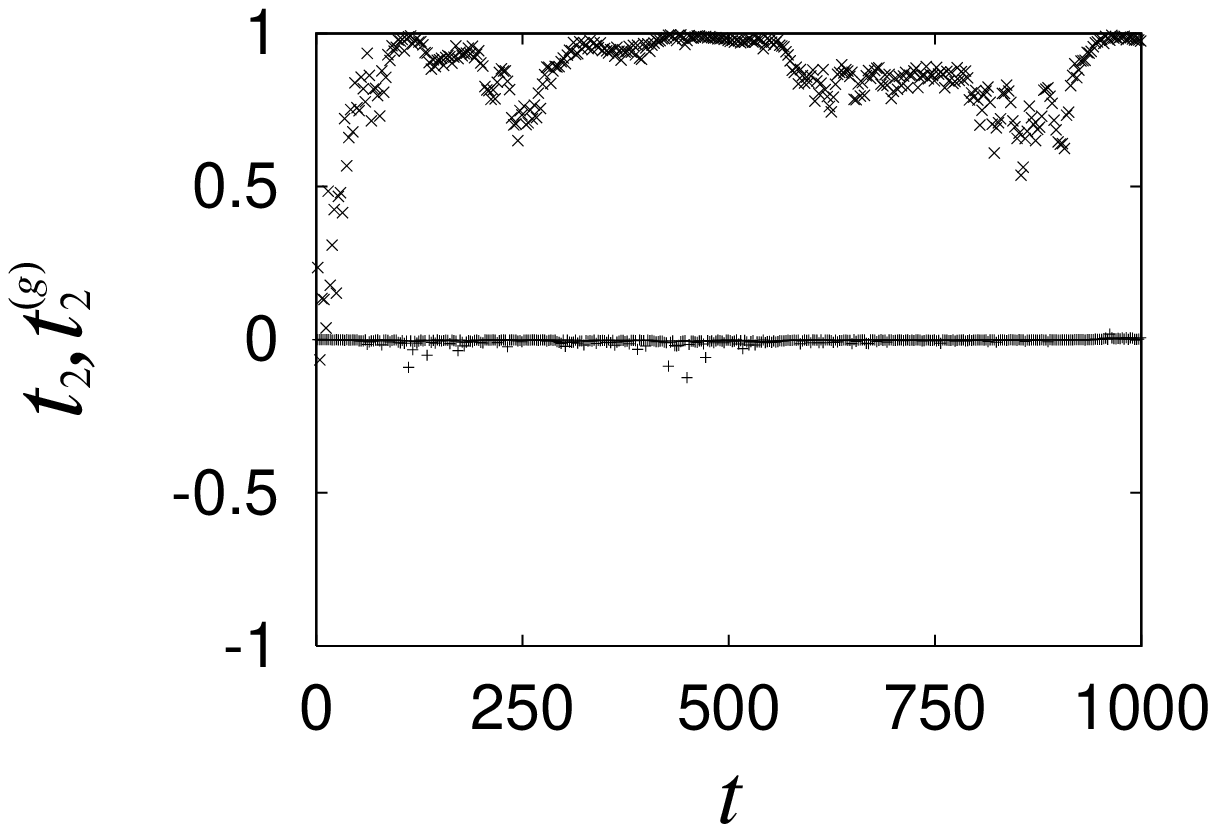}}
    \subfigure[$3$rd vector]{
      \includegraphics[width=5.5cm,clip,keepaspectratio]{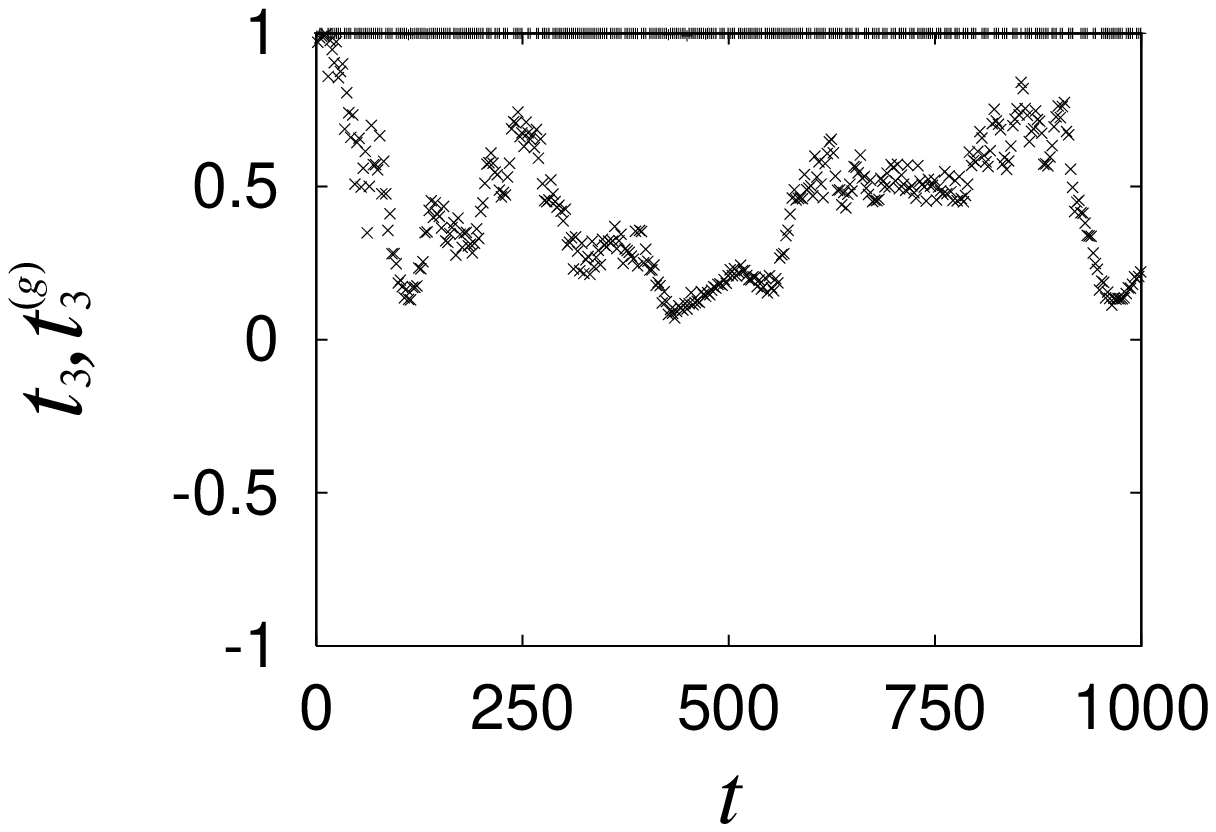}}\\
    \subfigure[$4$th vector]{
      \includegraphics[width=5.5cm,clip,keepaspectratio]{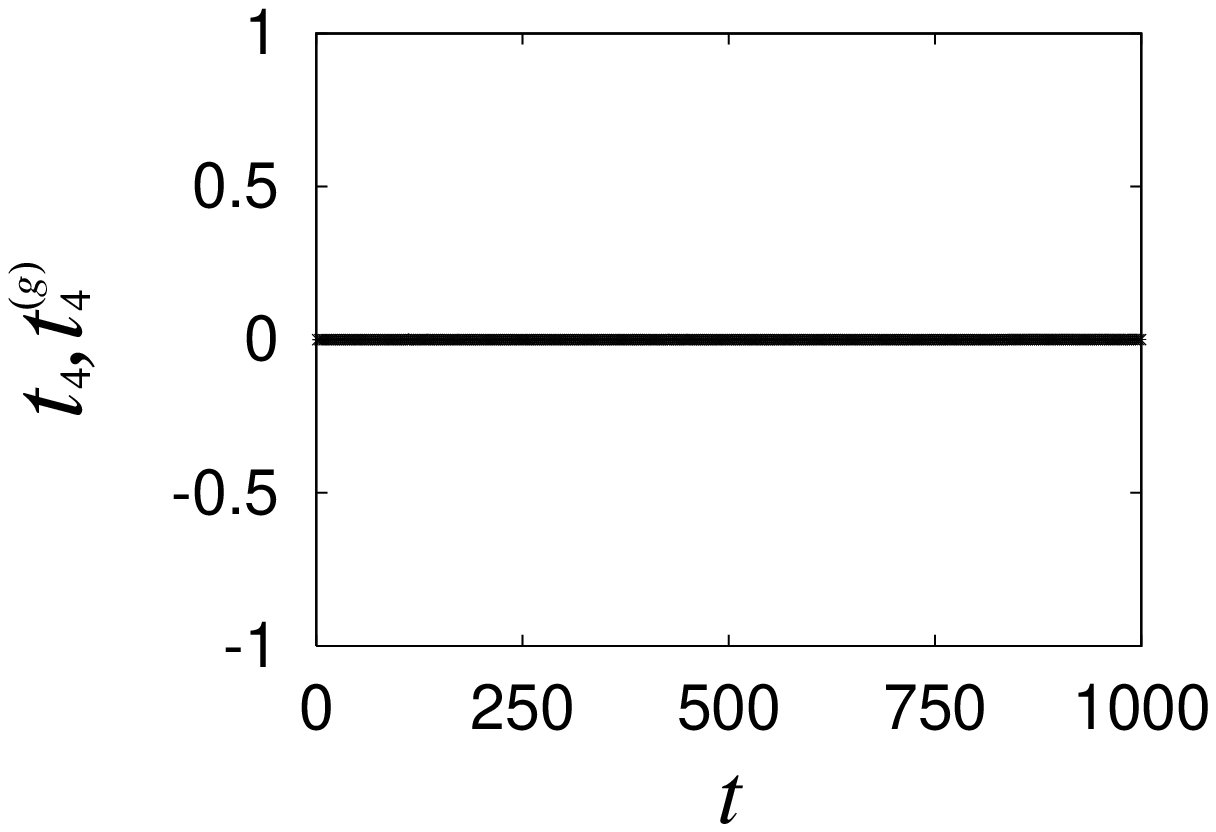}}
    \subfigure[$5$th vector]{
      \includegraphics[width=5.5cm,clip,keepaspectratio]{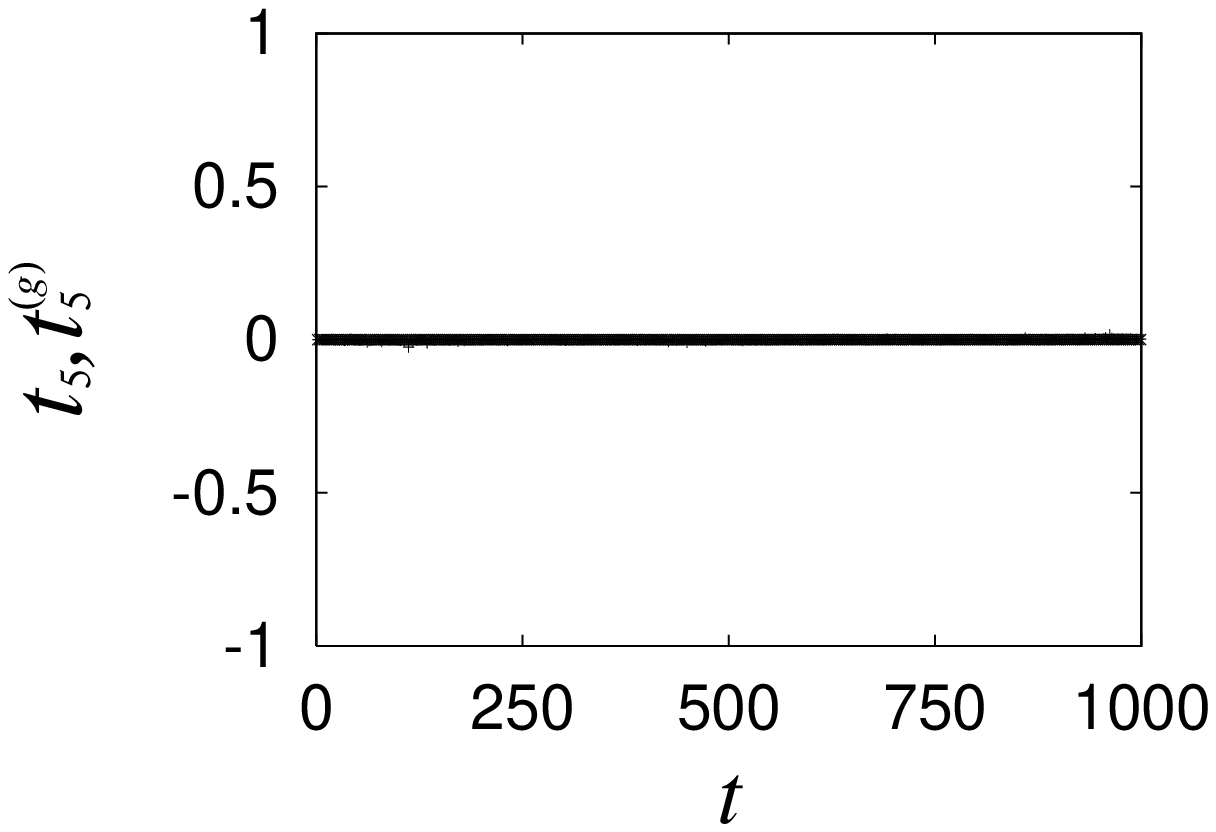}}
    \subfigure[$6$th vector]{
      \includegraphics[width=5.5cm,clip,keepaspectratio]{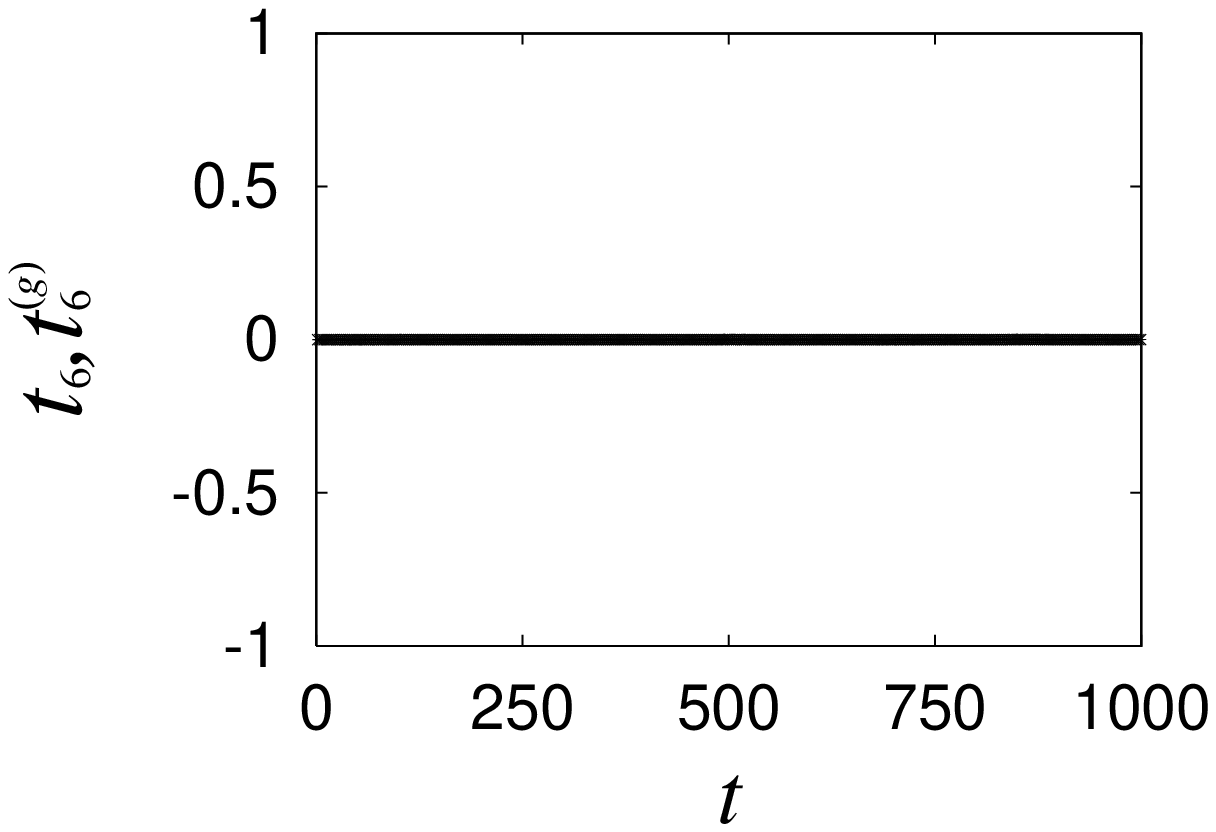}}
    \caption{Temporal evolutions of inner products
      between the normalized tangent vector of a Hamiltonian flow 
      and normalized Lyapunov vectors.
      The energy is set at $E=0.04$. 
      In (a),(b), and (c), straight lines
      are graphs of $t^{(g)}_{a}$
      %$\til{\bg}_J(\bV^{(g)}_{a},\bX_{\!K})$
      against the time parameter in the geometric method, 
      and broken curves are from the usual method, 
      providing the graphs of $t_{a}$.
      %$\til{\bg}_E(\bV_{a},\bX_{H})$. 
      The $1$st and $2$nd Lyapunov vectors $\bV_{1}^{(g)},\bV_{2}^{(g)}$ 
      are always orthogonal to the tangent direction to a Hamiltonian flow, 
      $\bX_{\!K}$, in the geometric method, but $\bV_{1},\bV_{2}$ are 
      not always orthogonal to $\bX_{\!H}$ in the usual method. 
      Moreover, the $3$rd Lyapunov vector always points to the
      direction of $\bX_{\!K}$ in the geometric method,
      but does not point to the direction of $\bX_{\!H}$ in the usual
      method.
      In (d),(e), and (f), only straight lines are drawn, 
      which are graphs from the both methods, but they coincide 
      with each other.}
    \label{fig:traj}
  \end{center}
\end{figure}

\begin{figure}[htbp]
 \begin{center}
    \subfigure[$1$st vector]{
      \includegraphics[width=5.5cm,clip,keepaspectratio]{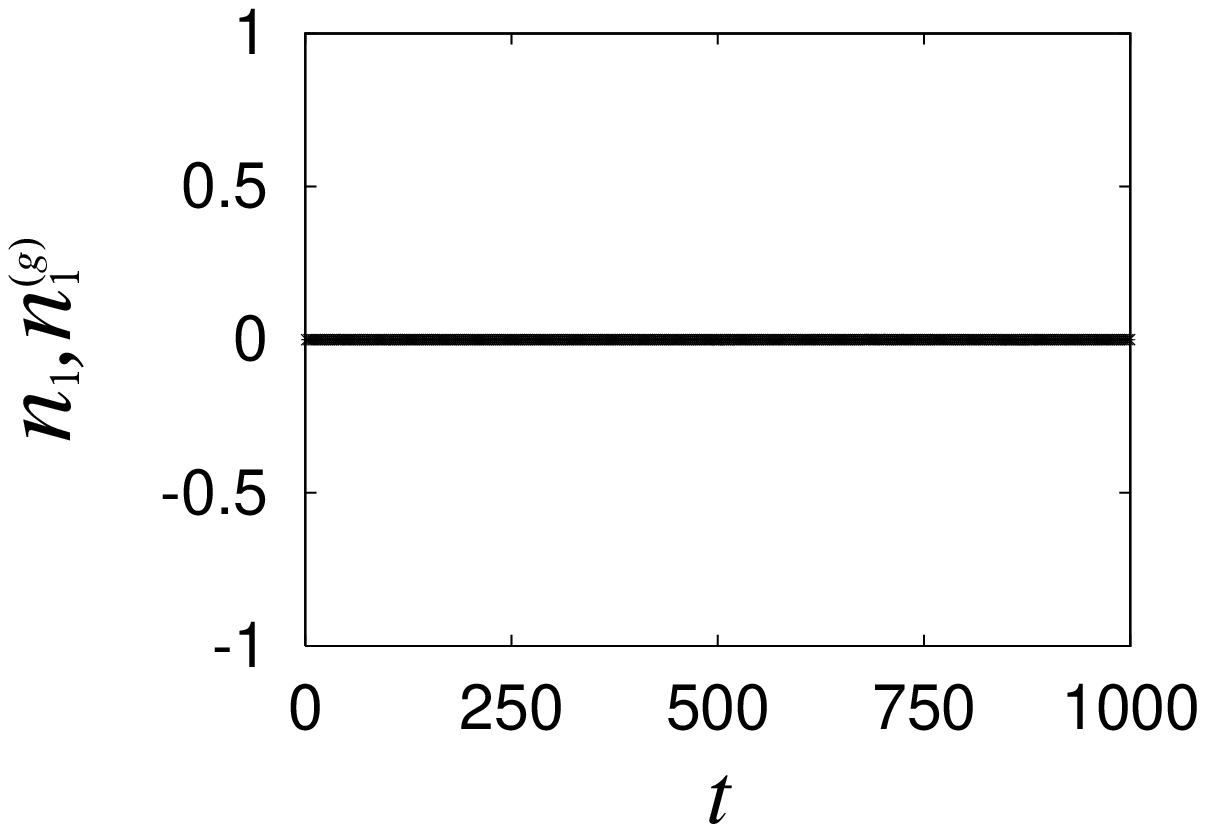}}
    \subfigure[$2$nd vector]{
      \includegraphics[width=5.5cm,clip,keepaspectratio]{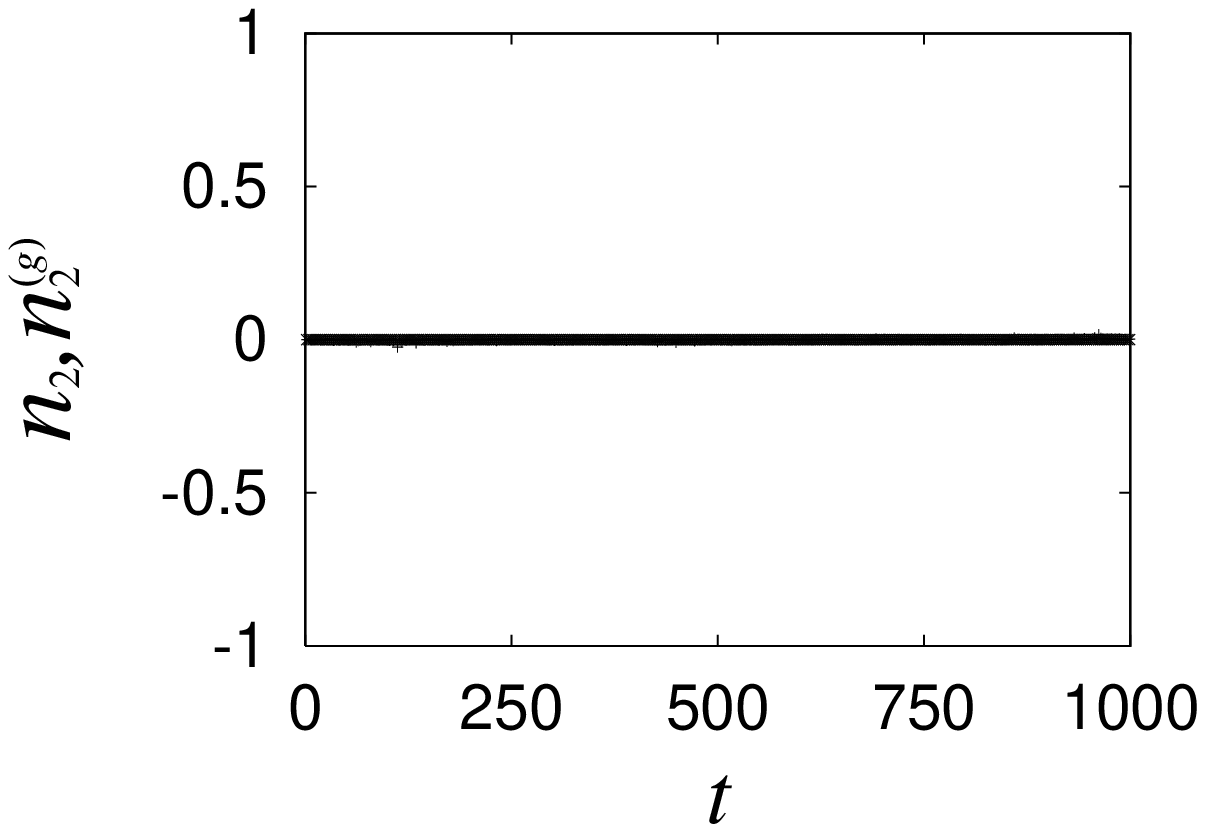}}
    \subfigure[$3$rd vector]{
      \includegraphics[width=5.5cm,clip,keepaspectratio]{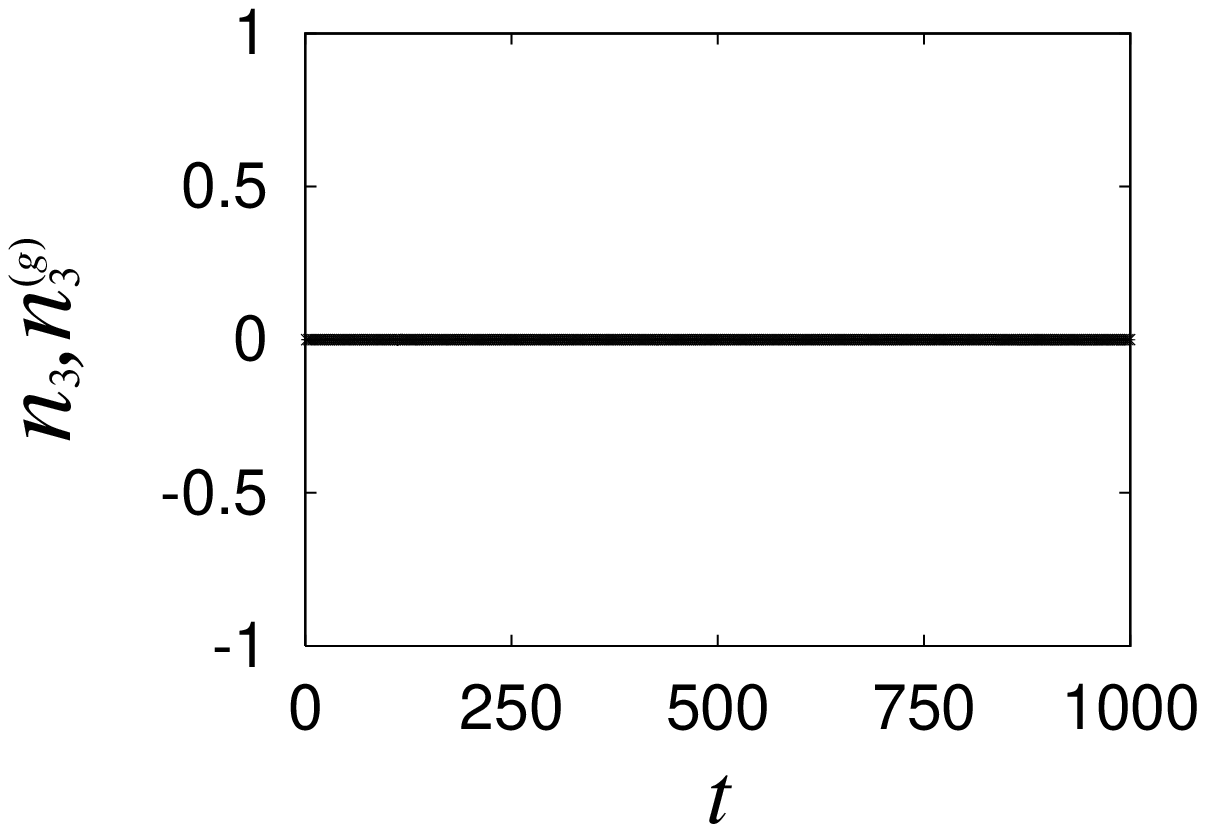}}\\
    \subfigure[$4$th vector]{
      \includegraphics[width=5.5cm,clip,keepaspectratio]{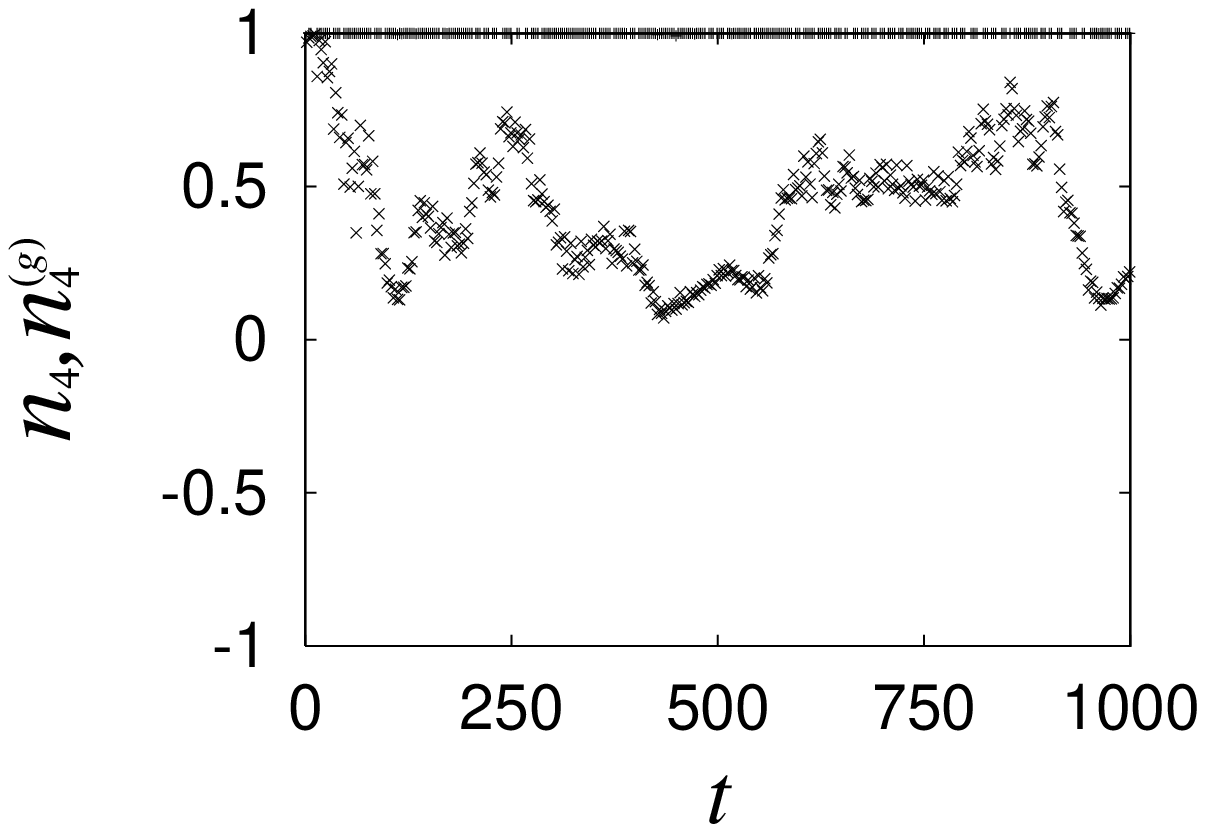}}
    \subfigure[$5$th vector]{
      \includegraphics[width=5.5cm,clip,keepaspectratio]{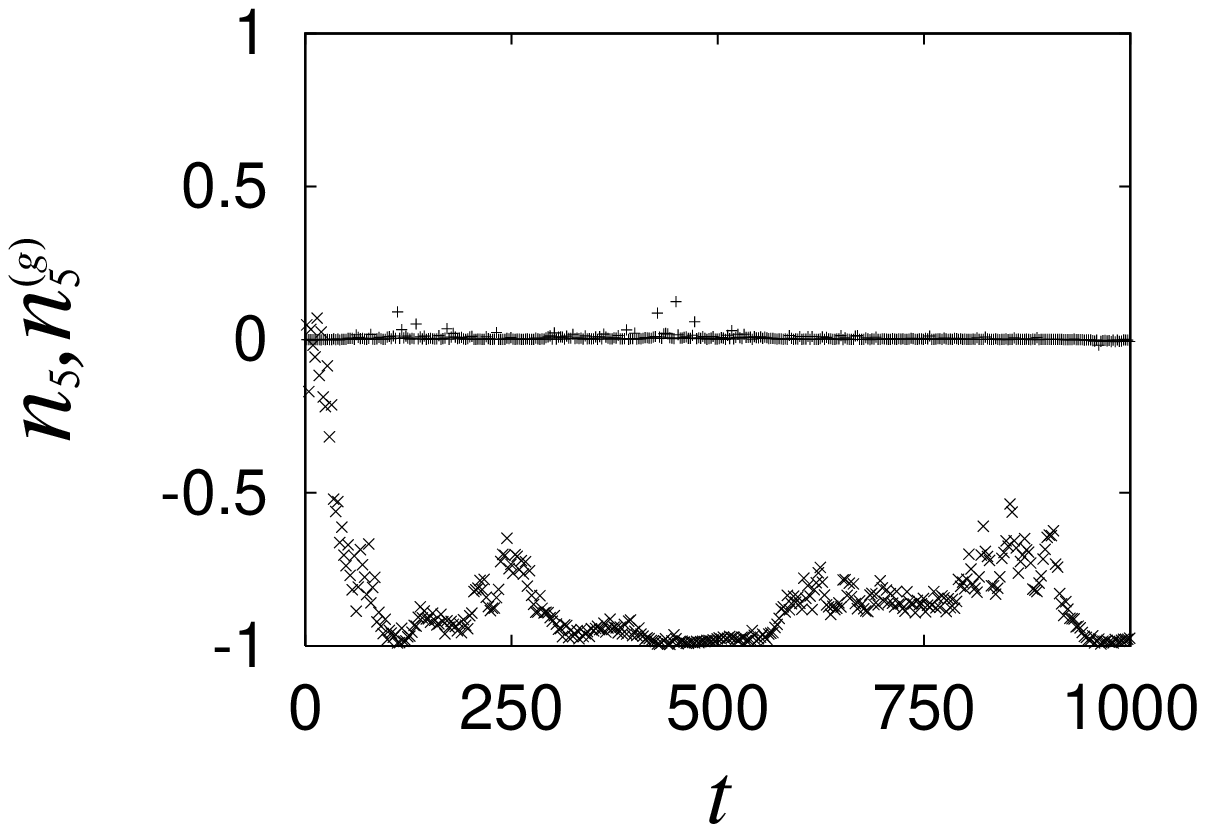}}
    \subfigure[$6$th vector]{
      \includegraphics[width=5.5cm,clip,keepaspectratio]{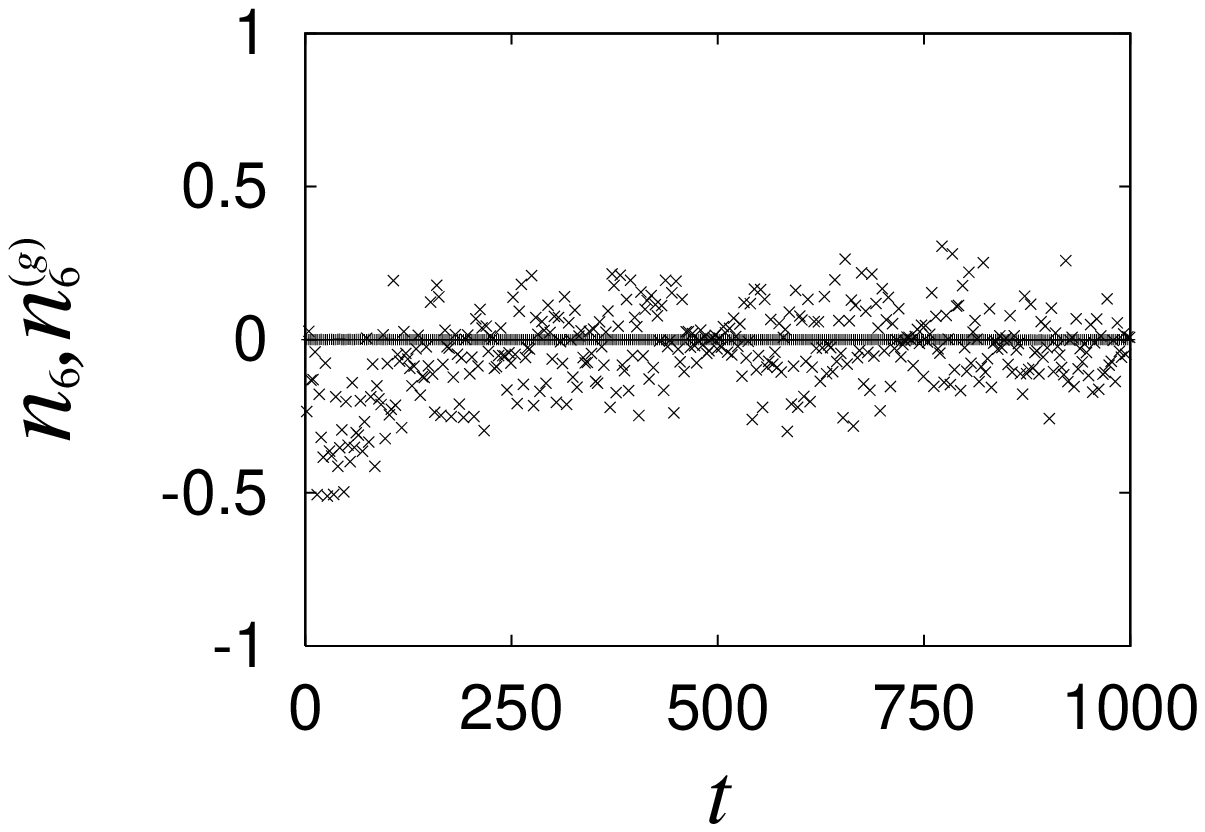}}
    \caption{Temporal evolutions of inner products between 
      the normalized gradient vector of the Hamiltonian function and
      the normalized Lyapunov vectors. 
      The energy is set at $E=0.04$. 
      In (d), (e), and (f), straight lines
      are graphs of $n^{(g)}_{a}$
      %$\til{\bg}_J(\bV^{(g)}_{a},\mbox{grad}K)$ 
      against the time parameter in the geometric method, 
      and broken curves are from the usual method, 
      providing the graph of $n_{a}$.
      %$\til{\bg}_E(\bV_{a},\mbox{grad}H)$. 
      The $4$th Lyapunov vector $\bV_{4}^{(g)}$ always points to the
      gradient direction of the Hamiltonian function $K$, 
      but $\bV_{4}$ does not always point to $\mbox{grad}H$ in the
      usual method. 
      Moreover, the $5$th and $6$th Lyapunov vectors
      $\bV_{5}^{(g)},\bV_{6}^{(g)}$ are always orthogonal to
      $\mbox{grad}K$ in the geometric method,  
      and $\bV_{5},\bV_{6}$ are not so to $\mbox{grad}H$ in the usual
      method. 
      In (a),(b), and (c), straight lines from the two methods
      are drawn, but each of them looks like a single line because of 
      coincidence.}
    \label{fig:grad}
  \end{center}
\end{figure}

From Figs.\ref{fig:traj} and \ref{fig:grad}, we will observe that 
the Lyapunov vectors calculated numerically in the geometric method 
satisfy the requirements stated in Sec.\ref{sec:introduction} and that 
the Lyapunov vectors calculated in the usual method have the property 
shown in Sec.\ref{sec:orthgonality-usual}.  
%By $\bV^{(g)}_{a}$ and $\bV_{a}$, $a=1,\cdots,6$, we denote the Lyapunov 
%vectors which are obtained in the geometric method and in 
%the usual method, respectively, to tell the difference between them. 

Figure \ref{fig:traj} provides temporal evolutions of inner products
between normalized Lyapunov vectors and the normalized tangent vector
to a Hamiltonian flow. 
The inner products both in the geometric method and in the 
usual method are denoted by $t^{(g)}_{a}$ and by $t_{a}$, 
respectively:
\begin{displaymath}
  t^{(g)}_{a} = \til{\bg}_{J} \left(
    \dfrac{\bV^{(g)}_{a}}{\norm{\bV^{(g)}_{a}}},
    \bX_{K} \right),
  \qquad
  t_{a} = \til{\bg}_{E} \left(
    \dfrac{\bV_{a}}{\norm{\bV_{a}}},
    \dfrac{\bX_{H}}{\norm{\bX_{H}}} \right).
\end{displaymath}
Figure \ref{fig:traj} shows that all the Lyapunov vectors except for 
$\bV^{(g)}_{3}$ in the geometric method are orthogonal to $\bX_{\!K}$, 
and that the normalized $\bV_3^{(g)}$ is equal to $\bX_{\!K}$. 
On the other hand, we observe also from Fig.\ref{fig:traj} that  
$\bV_{1}$ and $\bV_{2}$ in the usual method are not always orthogonal 
to $\bX_{\!H}$, and that the normalized $\bV_{3}$ does not equal 
$\bX_{\!H}/\Vert\bX_{\!H}\Vert$, either. 
In particular, we remark that $t_{2}$ takes values around unity
in opposition to our requirement.

Figure \ref{fig:grad} provides temporal evolutions of inner products
between normalized Lyapunov vectors and the normalized gradient vector 
of the Hamiltonian,
and the inner product both in the geometric method and in the usual method 
is denoted by the symbol $n^{(g)}_{a}$ and $n_{a}$, respectively:
\begin{displaymath}
  n^{(g)}_{a} = \til{\bg}_{J} \left(
    \dfrac{\bV^{(g)}_{a}}{\norm{\bV^{(g)}_{a}}},
    \gradK \right),
  \qquad
  n_{a} = \til{\bg}_{E} \left(
    \dfrac{\bV_{a}}{\norm{\bV_{a}}},
    \dfrac{\gradH}{\norm{\gradH}} \right).
\end{displaymath}
All the Lyapunov vectors except for $\bV^{(g)}_{4}$ are observed to be 
orthogonal to $\mbox{grad}K$, and $\bV^{(g)}_{4}$ to be collinear to 
$\mbox{grad}K$ in the geometric method, as is expected.  
On the other hand, the Lyapunov vectors $\bV_{5},\bV_{6}$ in the usual method 
are not always orthogonal to $\gradH$, 
and $\bV_{4}$ does not point to the direction of $\gradH$, either. 
In particular, $n_{5}$ in the usual method is far from vanishing, taking 
values around minus unity. 

These observations agree to what we expect from the theory 
described in Secs.\ref{sec:lyapunov-vectors} and 
\ref{sec:orthgonality-usual}. 
We note in conclusion that tiny fluctuations around straight lines 
in Figs.\ref{fig:traj} and \ref{fig:grad}, 
in particular, Figs.\ref{fig:traj}(b) and \ref{fig:grad}(e),
seem to be numerical errors due to the factor 
$1/2(E-V)$ included in the metric $g^{ij}$,  
Christoffel symbol $\Gamma^{i}_{jk}$ 
and the Riemann curvature tensor $R_{ijkl}$.

%%%%%%%%%%%%%%%%%%%%%%%%%%%%%%%%%%%%%%%%%%%%%%%%
% Section 6
%%%%%%%%%%%%%%%%%%%%%%%%%%%%%%%%%%%%%%%%%%%%%%%%
\section{Concluding Remarks}
\label{sec:summary}

In this paper, we have developed a new geometric method 
in Lyapunov analysis for natural Hamiltonian systems with $N$ 
degrees of freedom, which is set up on the cotangent bundle 
of a Riemannian manifold endowed with the Jacobi metric. 
In contrast with our geometric method, the old or already-known 
geometric method is established on the Riemannian manifold 
with the Jacobi metric.  According to that method, one 
brings Newton's equations of motion for a natural dynamical system 
into geodesic equations for the Jacobi metric and 
uses Jacobi equations, linearized geodesic equations, 
to analyze orbital instability of trajectories. 
However, the Jacobi equations are second-order differential 
equations, while Lyapunov exponents and vectors are defined 
through first-order differential equations. 
We then need first-order differential equation to apply 
Lyapunov analysis.  According to our method, the Jacobi equations 
are lifted from Riemannian manifolds to their cotangent bundles 
to take the form of first-order differential equations. 

When the geometric method is applied, a question arises as to 
whether Lyapunov exponents remain unchanged in their values or not, 
in comparison with those obtained in the usual method. 
As we have already pointed out, the linearized equations in both methods 
are different from each other and can not be transformed to each other 
through the parameter transformation $\rmd s=2(E-V(q))\rmd t$, 
while the equations of motion in both methods are transformed to each other 
through the same parameter transformation. 
However, the numerical computation has shown that the values of Lyapunov 
exponents coincide with each other, independently of the choice of 
methods applied, as far as the model system with $3$ degrees of 
freedom is taken.  
We guess that the Lyapunov exponents are long-term averaged values, 
so that they are independent of the choice of Lyapunov vectors along 
trajectories, while Lyapunov vectors depend on the choice of methods. 
As for the parameters of trajectories in the both method, 
we assume that the change of 
parameters must be subject to the condition $0<\rmd s/\rmd t<\infty$ 
along trajectories. 
On this account, we expect that Lyapunov exponents are independent of 
the choice of methods for calculation.  We will find indeed the 
coincidence of Lyapunov exponents in the both methods from numerical 
computations for other model systems. 
Further, observations made from the Lyapunov exponents are expected 
to be independent of the choice of methods. 
For instance, a characteristics of the graph of Lyapunov spectra 
$\lambda_{i}$ against $i/N$, $i=1,\cdots,N$ \cite{livi-87b,yamaguchi-98a}, 
which are observed in the usual method for a wide class of Hamiltonian 
systems having nearest neighbor interactions, 
will be found, in the geometric method as well, to be the same as that 
observed already in the usual method.  
We wish our geometric method may afford a fresh insight into 
the observation through Lyapunov vectors. 

In our geometric method developed in this article, 
we can choose Lyapunov vectors so as to satisfy the following
requirements:
(i) Lyapunov vectors except for $N$-th and $(N+1)$-th vectors
are always orthogonal to both the tangent direction to a trajectory
and the gradient direction of the Hamiltonian function,
(ii) $N$-th Lyapunov vector points to the tangent direction of 
the trajectory, $\bX_{\!K}$, and (iii) 
$(N+1)$-th Lyapunov vector points to the gradient direction 
of the Hamiltonian function, $\gradK$. 
Along with such Lyapunov vectors, we can analyze orbital instability 
of Hamiltonian flows in phase spaces without influence of 
the two marginal directions pointed by $\bX_{\!K}$ and $\gradK$ 
which have vanishing Lyapunov exponents, $\lambda_{N}=\lambda_{N+1}=0$. 
Moreover, the $N$-th and the $(N+1)$-th local Lyapunov exponents, 
which are averages of exponential growth rate in finite time, 
vanish on any time interval. 
The local Lyapunov exponents in the usual method are used, 
for instance, to distinguish nearly integrable 
systems from the others \cite{yamaguchi-96}.

%%%%%%%%%%%%%%%%%%%%%%%%%%%%%%%%%%%%%%%%%%%%%%%%%%%%%%%%%%%%%
In this article, we have considered the Hamiltonian function 
of the form $H(q,p)=\frac12\sum_{ij}\delta^{ij}p_{i}p_{j}+V(q)$ and 
developed the geometric method in Lyapunov analysis. 
However, the geometric method can be established for Hamiltonian 
functions of the form 
$H(q,p)=\frac12\sum_{ij}a^{ij}(q)p_{i}p_{j}+V(q)$, 
where $(a^{ij}(q))$ is the inverse of a metric tensor $(a_{ij}(q))$. 
In this case, the Jacobi metric is defined to be 
$g_{ij}(q)=2(E-V(q))a_{ij}(q)$, 
and geodesic equations for this metric 
\begin{displaymath}
  \dfracd{{}^2q^i}{s^2}+\Gamma^i_{jk}\dfracd{q^j}{s}\dfracd{q^k}{s}=0
\end{displaymath}
prove to be equivalent to Newton's equations of motion
\begin{displaymath}
  \dfracd{{}^2q^i}{t^2}+\christoffel{i}{j}{k}\dfracd{q^j}{t}\dfracd{q^k}{t}
=-a^{ij}\frac{\partial V}{\partial q^j}
\end{displaymath}
with the total energy fixed at $E$, where $\Gamma^i_{jk}$ and 
$\christoffel{i}{j}{k}$ are the Christoffel symbols formed from the 
metric $g_{ij}$ and $a_{ij}$, respectively, 
and $s$ is the length parameter for the Jacobi metric $g_{ij}$, 
which is related to the parameter $t$ by 
$\displaystyle{\dfracd{s}{t}}=2(E-V(q))$. 
The geometric method we have developed in  Lyapunov analysis of linearized 
Hamilton's equations of motion on the cotangent bundle is independent of 
the choice of the Riemannian metric chosen, 
so that the theorem stated in Sec.\ref{sec:lyapunov-vectors} holds also true 
in this case. Hence, we may find Lyapunov vectors which satisfy the 
requirements mentioned frequently. 
%%%%%%%%%%%%%%%%%%%%%%%%%%%%%%%%%%%%%%%%%%%%%%%%%%%%%%%%%%%%%

%%%%%%%%%%%%%%%%%%%%
%% acknowledgement
%%%%%%%%%%%%%%%%%%%%
\acknowledgements
The authors would like to thank Dr.~Y.~Uwano for valuable discussions,
and also the referee(s), whose comments helped them to brush up the
manuscript.
This work is supported by the Grand-In-Aid for Scientific Research of
the Ministry of Education, Culture, Sports, Science and Technology
of Japan (12750060 and 11640199).

%\newpage
%%%%%%%%%%%%%%%%%%%%
%% bibliography
%%%%%%%%%%%%%%%%%%%%

\appendix
%%%%%%%%%%%%%%%%%%%%%%%%%%%%%%%%%%%%%%%%%%%%
% Appendix A
%%%%%%%%%%%%%%%%%%%%%%%%%%%%%%%%%%%%%%%%%%%%
\section{Geometry of Cotangent Bundles}
\label{sec:appendixA}

Vector fields and Levi-Civita connection on a Riemannian 
manifold $M$ are lifted to the cotangent bundle $\TsM$,
and thereby the relation between geodesics on $M$ and 
geodesic flows on $\TsM$ will be made clear in geometric fashion.

\subsection{Lift of vector fields on $M$}

The cotangent bundle $\TsM$ is endowed with the standard 
one-form $\theta$, which is expressed locally as 
$\theta=p_{i}dx^{i}$. 
Note that the $\theta$ is defined globally on $\TsM$. 
This can be seen from the coordinate transformation 
on the non-empty intersection (\ref{eq:coordinates-transf}).
The exterior derivative of $\theta$, $\omega:=\rmd\theta$, 
is the standard symplectic form on $\TsM$. 

For vector fields on $M$, a way to lift them is not unique. 
A canonical way is given as follows: 
For $\bY\in\Xbar(M)$, the lifted vector field $\til{\bY}$ is 
defined through the conditions 
\begin{equation}
  \label{eq:lift}
  \pi_{*}\til{\bY}=\bY,\qquad  
  \mathcal{L}_{\til{\bY}} \theta = 0,
\end{equation}
where $\pi_*$ is the differential of the canonical projection $\pi$, 
and $\mathcal{L}$ denotes the Lie derivation.
For $\bY=Y^{i}\partial_{i}$, a straightforward calculation shows 
that the $\til{\bY}$ is put in the form
\begin{equation}
  \label{eq:lift-vector}
  \til{\bY} = Y^{i} \dfracp{}{x^{i}} 
   - p_{j} \dfracp{Y^{j}}{x^{i}} \dfracp{}{p_{i}} .
\end{equation}
Furthermore, owing to Cartan's formula, 
$\mathcal{L}_{\til{\bY}} \theta
  = \rmd (\iota(\til{\bY}) \theta) + \iota(\til{\bY}) 
  \rmd \theta $, along with 
$\iota(\til{\bY})\theta=\theta(\til{\bY})$,
the latter of the conditions (\ref{eq:lift}) implies that 
$-d(\theta(\til{\bY}))=\iota(\til{\bY})\omega$, 
which then shows that the $\til{\bY}$ becomes the Hamiltonian 
vector field associated with $F:=\theta(\til{\bY})=p_{i}Y^{i}$.  
Thus one has 
\begin{equation}
  \label{eq:Ham-v-F}
  \til{\bY} = \bX_{\!F} 
  =\dfracp{F}{p_{i}} \dfracp{}{x^{i}}
  - \dfracp{F}{x_{i}}\dfracp{}{p^{i}}.
\end{equation}
With respect to the adapted frame, 
the canonical lift $\til{\bY}$ takes the form
\begin{equation}
  \label{eq:lift-vector-adapted-frame}
  \til{\bY} = Y^{i} D_{i} - p_{j} \nabla_{i} Y^{j} D_{\up{i}},
\end{equation}
where
\begin{displaymath}
  \nabla_{i} Y^{j} = \dfracp{Y^{j}}{x^{i}} + \Gamma^{j}_{ik} Y^{k}.
\end{displaymath}

In addition to the canonical lift, one can define another lift; 
for a vector field $\bY=Y^{i}\partial_{i}$ on $M$, the horizontal lift
of $\bY$ is given on $\TsM$ by 
\begin{equation}
   \label{eq:hor-lift}
  \til{\bY}^{h} = Y^{i} D_{i}.
\end{equation}
From the transformation rule (\ref{eq:f-transf}), 
the horizontal lift is shown to be defined independently of 
the choice of adapted frames. 
\par
A curve $x(t)$ in $M$ is also lifted horizontally; a curve 
$\til{x}^{h}(t)$ in $\TsM$ is called a horizontal lift of $x(t)$,  
if $\pi(\til{x}^{h}(t))=x(t)$ and if the tangent vector 
to $\til{x}^{h}(t)$ is horizontal.  
To give an example of horizontal lifts of curves, we consider 
a geodesic $x(s)$ with $s$ the arc length parameter.  
Let $\bxi(s)$ denote its tangent vector and let
$p_{i}(s)=g_{ij}\xi^{j}(s)$.  
Then a curve $(x(s),p(s))$ in the cotangent bundle $\TsM$ is shown 
to be a horizontal curve. 
In fact, differentiation of $(x(s),p(s))$ with respect to $s$ 
along with the geodesic equation for $x(s)$ provides 
\begin{equation}
  \label{eq:h-lift-geo}
  \dfrac{\rmd}{\rmd s}(x^{i}(s),p_{i}(s))
  = (\xi^{i}, \Gamma^{k}_{ij} p_{k} \xi^{j})
  = \xi^{i}(s)D_{i}, 
\end{equation}
as is wanted. From Eq.(\ref{eq:h-lift-geo}) along with 
$\xi^{i}(s)=g^{i\ell}p_{\ell}(s)$, we observe that 
the curve $(x(s),p(s))$ is a geodesic flow, 
an integral curve of $\bX_{\!K}$ (see (\ref{eq:geod-spray})).  
\par
Now we assume that $\bxi$ is a tangent vector field to 
a congruence of geodesics in $M$.  According to 
(\ref{eq:hor-lift}), we can define the horizontal lift 
$\til{\bxi}^h$ on $T^*M$.  With the restriction 
$p_i=g_{ij}\xi^j(x)$ imposed, the Hamiltonian vector field 
$\bX_{\!K}$ becomes equal to the horizontal lift $\til{\bxi}^h$. 
Hence, a congruence of geodesics in $M$ is lifted to a family 
of geodesic flows in $T^*M$ along with  $\til{\bxi}^h=
\bX_{\!K}$. 
\par
We proceed to discuss lifts of geodesic deviations. 
Let $\bY(s)$ be a vector field defined along the geodesic $x(s)$. 
We define a vector field $\bX(s)$ along a geodesic flow 
$(x(s),p(s))$ with $p_i(s)=g_{ij}\xi^j(s)$, by 
\begin{equation}
  \label{eq:lifted-Y}
  \bX = Y^{i} D_{i} + g_{ij}(\nabla_{\bxi}\bY)^{j} D_{\up{i}}. 
\end{equation}
We note here that the $\bX(s)$ is defined independently of 
the choice of adapted frames. 
If $\bX(s)$ satisfies the lifted Jacobi equation 
(\ref{eq:deviation-geo2}), then 
the $\bY(s)$ should be a Jacobi field. 
Conversely, for a given Jacobi field $\bY(s)$ defined along 
a geodesic $x(s)$, we can form a lifted vector field $\bX(s)$  
according to (\ref{eq:lifted-Y}), which is defined along a geodesic 
flow $P(s)=(x(s),p(s))$ with $p_{i}(s)=g_{ij}\xi^{j}(s)$. 
Then $\bX(s)$ solves Eq.(\ref{eq:deviation-geo2}). 

\subsection{Killing vector fields}
\label{sec:Killing}

We now wish to investigate the relation between the canonical 
lift (\ref{eq:Ham-v-F}) and the lift (\ref{eq:lifted-Y}), 
where $\bxi$ is viewed as the tangent vector field to a 
congruence of geodesics in $M$. 
To this end, we first consider symmetry of our Hamiltonian system 
with the Hamiltonian function $K$. 
We assume here that for a vector field $\bY$ on $M$ 
the function $F=\theta(\bY)=Y^ip_i$ is a constant of motion; 
$\bX_{\!K}(F)=-\{K,F\}=0$, 
where $\{\cdot,\cdot\}$ denotes Poisson bracket.  
Then one obtains 
$[\bX_{\!K},\bX_{\!F}]=-\bX_{\!\{K,F\}}=0$. 
This implies that $\til{\bY}=\bX_{\!F}$ satisfies the linearized 
equation (\ref{eq:deviation-geo2}) along any geodesic flow. 
On the other hand, the condition $\bX_{\!K}(F)=0$ holds, 
if and only if $\bY$ is a Killing vector field, an infinitesimal 
isometry, {\it i.e.}, ${\cal L}_{\bY}\bg=0$, as is easily seen. 
It is well known that every Killing vector field satisfies the 
Jacobi equation along any geodesic. 
\par
Now we assume further that we are given the tangent vector field 
$\bxi$ to a congruence of geodesics in $M$. 
If restricted on a subspace $L$ determined by $p_i=g_{ij}\xi^j$ in 
$T^*M$, the canonical lift $\til{\bY}$ of a Killing vector 
field $\bY$ is expressed as 
\begin{equation}
  \til{\bY}|_{L} = Y^iD_i-g_{jk}\xi^k\nabla_i Y^j D_{\up{i}} 
  = Y^iD_i + g_{ij}(\nabla_{\bxi}\bY)^j D_{\up{i}}, 
\end{equation}
where use has been made of the formula that 
\begin{equation}
  g^{ij}\nabla_i Y^k + g^{ik}\nabla_i Y^j = 0, 
\end{equation}
which is a necessary and sufficient condition for $\bY$ 
to be a Killing vector field. 
Thus we have found that if $\bY$ is a Killing vector on $M$, 
and if the canonical lift $\til{\bY}$ is restricted to $L$ 
determined by $p_i=g_{ij}\xi^j$, then $\til{\bY}|L$ is equal 
to the lift (\ref{eq:lifted-Y}) with $\bxi$ the tangent vector 
field to a congruence of geodesics.

\subsection{Levi-Civita connection of $\TsM$}
\label{sec:lift-chistoffel}

The Levi-Civita connection $\widetilde{\nabla}$ is defined on the 
cotangent bundle $\TsM$ through the Sasaki metric $\til{\bg}$. 
We denote the Christoffel symbols for this connection 
by $\hat{\Gamma}^A_{BC}$; 
\begin{displaymath}
   \widetilde{\nabla}_{\partial_{B}} \partial_{C} =
   \hat{\Gamma}^{A}_{BC}\partial_{A},
\end{displaymath}
where Roman capital indices run from $1$ to $2m$ and 
$\partial_{A}$ are the standard frame;
\begin{displaymath}
 \partial_{i} = \dfracp{}{x^{i}} , \qquad
 \partial_{\up{i}} = \dfracp{}{p_{i}} ,\qquad
 i=1,\cdots,m. 
\end{displaymath}
The Christoffel symbols are given, as usual, by 
\begin{displaymath}
  \hat{\Gamma}^{A}_{BC}
  = \dfrac{1}{2} \hat{g}^{AD}
  \left(
    \partial_{B} \hat{g}_{CD} + \partial_{C} \hat{g}_{DB}
    - \partial_{D} \hat{g}_{BC}
  \right) ,
\end{displaymath}
where $\hat{g}_{AB}$ are components of $\til{\bg}$; 
$\hat{g}_{AB}=\til{\bg}(\partial_{A},\partial_{B})$.
We denote the coefficients of the connection $\widetilde{\nabla}$ 
with respect to the adapted frame by 
$\widetilde{\Gamma}^{\alpha}_{\beta\gamma}$;
\begin{displaymath}
   \widetilde{\nabla}_{D_{\beta}}D_{\gamma}=
   \widetilde{\Gamma}^{\alpha}_{\beta\gamma}D_{\alpha}, 
\end{displaymath}
where Greek indices also run from $1$ to $2m$, 
but indicating that they are indices for the adapted frame.

Let the functions $\Omega_{\beta\gamma}^{\hspace*{1.0em}\alpha}$ 
be defined by 
\begin{displaymath}
  [ D_{\beta}, D_{\gamma} ] =
  \Omega_{\beta\gamma}^{\hspace*{1.0em}\alpha} D_{\alpha}. 
\end{displaymath}
Then the torsion-free condition for $\widetilde{\nabla}$ is 
put in the form 
\begin{displaymath}
  \til{\Gamma}^{\alpha}_{\beta\gamma} 
  - \til{\Gamma}^{\alpha}_{\gamma\beta}
  = \Omega_{\beta\gamma}^{\hspace*{1.0em}\alpha}.
\end{displaymath}
A straightforward calculation yields 
$\Omega_{\beta\gamma}^{\hspace*{1.0em}\alpha}$ as follows:
\begin{displaymath}
  \begin{split}
  [D_i,D_j] & =p_{\ell} R_{ijk}^{\hspace*{1.2em}\ell}D_{\up{k}},\quad
  [D_i,D_{\up{j}}]  =-\Gamma^j_{ik}D_{\up{k}},\\[2mm]
  [D_{\up{i}},D_j] & =\Gamma^i_{jk}D_{\up{k}},\qquad 
  [D_{\up{i}},D_{\up{j}}]  =0.
  \end{split}
\end{displaymath}

We are to write out $\til{\Gamma}^{\alpha}_{\beta\gamma}$ 
in terms of $\til{g}_{\alpha\beta}$ and 
$\Omega_{\beta\gamma}^{\hspace*{1.0em}\alpha}$, where 
$\til{g}_{\alpha\beta}=\til{\bg}(D_{\alpha},D_{\beta})$, 
the components of $\til{\bg}$ with respect to the adapted frame. 
The covariant derivative of the metric $\til{\bg}$ must 
vanish for all vector fields $\bX$ on $\TsM$; 
$\til{\nabla}_{\bX} \til{\bg}=0$, 
so that one has 
\begin{displaymath}
  D_{\beta} \til{g}_{\gamma\delta}
  - \til{\Gamma}^{\varepsilon}_{\beta\gamma} 
  \til{g}_{\varepsilon\delta}
  - \til{\Gamma}^{\varepsilon}_{\beta\delta}
  \til{g}_{\gamma\varepsilon}
  = 0.
\end{displaymath}
Further calculation provides 
\begin{displaymath}
  \begin{split}
    D_{\beta} \til{g}_{\gamma\delta}
    + D_{\gamma} \til{g}_{\delta\beta}
    - D_{\delta} \til{g}_{\beta\gamma}
    & = \left( 
      \til{\Gamma}^{\epsilon}_{\beta\gamma}
      + \til{\Gamma}^{\epsilon}_{\gamma\beta} 
    \right) \til{g}_{\epsilon\delta}
    + \left(
      \til{\Gamma}^{\epsilon}_{\beta\delta} 
      - \til{\Gamma}^{\epsilon}_{\delta\beta} 
    \right) \til{g}_{\epsilon\gamma}
    + \left(
      \til{\Gamma}^{\epsilon}_{\gamma\delta} 
      - \til{\Gamma}^{\epsilon}_{\delta\gamma}
    \right) \til{g}_{\epsilon\beta} \\
    & = \left( 
      2 \til{\Gamma}^{\epsilon}_{\beta\gamma}
      - \Omega_{\beta\gamma}^{\ \ \ \epsilon}
    \right) \til{g}_{\epsilon\delta}
    + \Omega_{\beta\delta}^{\ \ \ \epsilon} 
    \til{g}_{\epsilon\gamma}
    + \Omega_{\gamma\delta}^{\ \ \ \epsilon} 
    \til{g}_{\epsilon\beta} ,
  \end{split}
\end{displaymath}
which results in 
\begin{displaymath}
  \til{\Gamma}^{\alpha}_{\beta\gamma} 
  = \dfrac{1}{2} \til{g}^{\alpha\delta} \left(
    D_{\beta} \til{g}_{\gamma\delta}
    + D_{\gamma} \til{g}_{\delta\beta}
    - D_{\delta} \til{g}_{\beta\gamma}
  \right)
  + \dfrac{1}{2} \left(
    \Omega_{\beta\gamma}^{\ \ \ \alpha} 
    + \Omega^{\alpha}_{\ \beta\gamma}
    + \Omega^{\alpha}_{\ \gamma\beta}
  \right) ,
\end{displaymath}
where
\begin{displaymath}
  \Omega^{\alpha}_{\ \beta\gamma}
  =  \til{g}^{\alpha\delta} \Omega_{\delta\beta}^{\ \ \ \epsilon
}
  \til{g}_{\epsilon\gamma} .
\end{displaymath}
A straightforward calculation shows that 
the components $\til{\Gamma}^{\alpha}_{\beta\gamma}$ 
are given by 
\begin{equation}
  \label{eq:lift-Christoffel}
  \begin{split}
    & 
    \til{\Gamma}^{i}_{jk} = \Gamma^{i}_{jk}, \quad
    \til{\Gamma}^{i}_{j\up{k}} 
    = - \frac{1}{2} R_{j}^{\hspace*{0.3em}ik\ell} p_{\ell}, 
    \quad
    \til{\Gamma}^{i}_{\up{j}k} 
    = - \frac{1}{2} R_{k}^{\hspace*{0.4em}ij\ell} p_{\ell}, 
    \quad
    \til{\Gamma}^{i}_{\up{j}\up{k}} = 0, \\
    & 
    \til{\Gamma}^{\up{i}}_{jk} 
    = \frac{1}{2} R_{jki}^{\hspace*{1.2em}\ell} p_{\ell}, 
    \quad
    \til{\Gamma}^{\up{i}}_{j\up{k}} 
    = - \Gamma^{k}_{ij} , \quad
    \til{\Gamma}^{\up{i}}_{\up{j}k} 
    = 0 , \quad
    \til{\Gamma}^{\up{i}}_{\up{j}\up{k}} = 0.
  \end{split}
\end{equation}
Covariant derivatives of vector fields are then expressed, 
in terms of these coefficients, as 
\begin{equation}
  \label{eq:covariant-derivative}
  \til{\nabla}_{\bX_{1}} \bX_{2}
  = \left[
    X_{1}^{\beta} D_{\beta} X_{2}^{\alpha}
    + \til{\Gamma}^{\alpha}_{\beta\gamma} X_{1}^{\beta}
    X_{2}^{\gamma} 
  \right] D_{\alpha},
\end{equation}
where $(X_{1}^{\alpha})$ and $(X_{2}^{\alpha})$ are components of
$\bX_{1}$ and $\bX_{2}$ with respect to $D_{\alpha}$, respectively.
In particular, the covariant derivative of 
$\bX=X^{i}D_{i}+X^{\up{i}}D_{\up{i}}$ with respect to the horizontal 
lift $\til{\bxi}^{h}=\xi^i(s)D_i$ along a geodesic flow as a horizontal 
lift of a geodesic takes the form 
\begin{equation}
  \label{eq:codev-xi}
  \begin{split}
    (\til{\nabla}_{\til{\bxi}^{h}} {\bX})^{i}
    & = \dfracd{X^{i}}{s} + \Gamma^{i}_{kj} \xi^{k} X^{j}
    - \dfrac{1}{2} R_{k}^{\hspace*{0.3em}ij\ell} p_{\ell} 
    \xi^{k} X^{\up{j}} , \\
    (\til{\nabla}_{\til{\bxi}^{h}} {\bX})^{\up{i}}
    & = \dfracd{X^{\up{i}}}{s} - \Gamma^{j}_{ik} \xi^{k} 
    X^{\up{j}}
    + \dfrac{1}{2} R_{kji}^{\hspace*{1.0em}\ell} p_{\ell} 
    \xi^{k}  X^{j} .
  \end{split}
\end{equation}
If $\bX=\til{\bxi}^{h}$, these equations give rise to 
\begin{displaymath}
  \til{\nabla}_{\til{\bxi}^h} \til{\bxi}^h = 0,
\end{displaymath}
which implies that the horizontal lift $(x(s),p(s))$ 
of a geodesic $x(s)$ on $M$ is also a geodesic on 
$\TsM$ with respect to the lifted metric $\til{\bg}$. 
We note here that the arc length parameter $\sigma$ with respect 
to $\til{\bg}$ reduces to the arc length parameter $s$, if 
the curve is horizontal. 

%%%%%%%%%%%%%%%%%%%%%%%%%%%%%%%%%%%%%%%%%%%%
% Appendix B
%%%%%%%%%%%%%%%%%%%%%%%%%%%%%%%%%%%%%%%%%%%%
\section{Symplectic Implicit Runge-Kutta method}
\label{sec:appendixB}

\begin{table}[hptb]
  \begin{center}
    \caption{Kuntzmann \& Butcher method, order $6$.
      The upper right block is the matrix $(a_{ij})$,
      the lower raw is the vector $(b_{i})$
      and the left column is the vector $(c_{i})$.}
    \renewcommand{\arraystretch}{2.8}
    \begin{tabular}{c | ccc}
      $\dfrac{1}{2} - \dfrac{\sqrt{15}}{10}\ $ 
      & $\dfrac{5}{36}$ 
      & $\dfrac{2}{9} - \dfrac{\sqrt{15}}{15}\ $ 
      & $\dfrac{5}{36} - \dfrac{\sqrt{15}}{30}\ $ \\
      $\dfrac{1}{2}$ 
      & $\dfrac{5}{36} + \dfrac{\sqrt{15}}{24}\ $ 
      & $\dfrac{2}{9}$ 
      & $\dfrac{5}{36} - \dfrac{\sqrt{15}}{24}\ $ \\
      $\dfrac{1}{2} + \dfrac{\sqrt{15}}{10}\ $ 
      & $\dfrac{5}{36} + \dfrac{\sqrt{15}}{30}\ $ 
      & $\dfrac{2}{9} + \dfrac{\sqrt{15}}{15}\ $ 
      & $\dfrac{5}{36}$ \\
      \hline
      & $\dfrac{5}{18}$ & $\dfrac{4}{9}$ & $\dfrac{5}{18}$ 
    \end{tabular}
    \label{tab:kb6}
  \end{center}
\end{table}

Suppose we are given a dynamical system in ${\bR}^{\ell}$
\begin{equation}
  \label{eq:eq1}
  \dfracd{x}{t}(t) = f(x,t).
\end{equation}
Numerical integration of this equation is performed through 
discretizing it with time slice $h$.
The $s$-stage Runge-Kutta method for integration is given by 
\begin{displaymath}
  \begin{split}
    x' & = x + h \sum_{i=1}^{s} b_{i} k_{i} \\
    k_{i} & = f( x + h \sum_{j=1}^{s} a_{ij} k_{j}, t + c_{i} h), 
    \quad i=1,\cdots,s,
  \end{split}
\end{displaymath}
where $(x,t)$ goes to $(x',t+h)$ after one step,
and $a_{ij},b_{i}$ and $c_{i}$ are real constants
with $\sum_{i=1}^{s} c_{i} =1$.
Note that the second of the above equations defines implicitly $k_i$. 
The $3$-stage Runge-Kutta method,
namely the $6$-th order Kuntzmann \& Butcher method,
is defined as in Table \ref{tab:kb6}.


\begin{thebibliography}{10}

\bibitem{oseledec-68}
V.~I.~Oseledec,
Trans.~Moscow~Math.~Soc. {\bf 19}, 197 (1968).

\bibitem{benettin-76}
G.~Benettin, L.~Galgani and J.-M.~Strelcyn,
Phys.~Rev.~A {\bf 14}, 2338 (1976).

\bibitem{firpo-98}
M.-C.~Firpo,
Phys.~Rev.~E {\bf 57}, 6599 (1998).

\bibitem{butera-87}
P.~Butera and G.~Caravati,
Phys.~Rev.~A {\bf 36}, 962 (1987).

\bibitem{mutschke-93}
G.~Mutschke and U.~Bahr,
Physica~D {\bf 69}, 302 (1993).

\bibitem{konishi-92}
T.~Konishi and K.~Kaneko,
J.~Phys.~A:~Math.~Gen. {\bf 25}, 6283 (1992). 

\bibitem{sasa-00}
S.~Sasa,
nlin.CD/0006042.

\bibitem{pettini-93}
M.~Pettini,
Phys.~Rev.~E {\bf 47}, 828 (1993).

\bibitem{casetti-95}
L.~Casetti, R.~Livi and M.~Pettini,
Phys.~Rev.~Lett. {\bf 74}, 375 (1995).

\bibitem{casetti-96}
L.~Casetti, C.~Clementi and M.~Pettini,
Phys.~Rev.~E {\bf 54}, 5969 (1996).

\bibitem{casetti-00}
L.~Casetti, M.~Pettini, and E.~G.~D.~Cohen,
Phys.~Rep. {\bf 337}, 237 (2000).

\bibitem{whittaker-37}
E.~T.~Whittaker,
{\it Analytical Dynamics of Particles and Rigid Bodies}
(Cambridge University Press, 1937).

\bibitem{ong-75}
C.~P.~Ong,
Adv.~Math.~ {\bf 15}, 269 (1975).

\bibitem{yano-73}
K.~Yano and S.~Ishihara,
{\it Tangent and Cotangent Bundles}
(Marcel Dekker Inc., New York, 1973).

\bibitem{sasaki-58}
S.~Sasaki,
T\^ohoku~Math.~J. {\bf 10}, 338 (1958).

\bibitem{yoshida-90}
H.~Yoshida,
Phys.~Lett.~A {\bf 150}, 262 (1990).

\bibitem{casetti-95b}
L.~Casetti,
Physica Scripta {\bf 51}, 29 (1995).

\bibitem{hairer-93}
E.~Hairer, S.~P. No\hspace*{-0.5em}/rsett, and G.~Wanner,
{\it Solving ordinary differential equations I}
(Springer-Verlag, Berlin, second revised edition, 1993).

\bibitem{cerruti-sola-97}
M.~Cerruti-Sola and R.~Franzosi and M.~Pettini,
Phys.~Rev.~E {\bf 56}, 4872 (1997).

\bibitem{livi-87b}
R.~Livi, A.~Politi, S.~Ruffo, and A.~Vulpiani,
J.~Stat.~Phys. {\bf 46}, 147 (1987).

\bibitem{yamaguchi-98a}
Y.~Y.~Yamaguchi,
J.~Phys.~A {\bf 31}, 195 (1998).

\bibitem{yamaguchi-96}
Y.~Y.~Yamaguchi,
Prog.~Theor.~Phys. {\bf 95}, 717 (1996).

\end{thebibliography}
\end{document}